\documentclass[aip,graphicx]{revtex4-1}
\usepackage{mathtools}

\usepackage{ulem}
\usepackage{amsmath,amssymb,graphicx}
\usepackage[usenames]{color}
\usepackage{multirow}
\usepackage{array}
\usepackage{longtable}
\usepackage{graphicx}
\usepackage{url}
\usepackage{graphicx,epstopdf}
\fontsize{11}{11}
\newcommand\comment[1]{}

\newcommand\beq {\begin{equation}}
\newcommand\eeq {\end{equation}}

\newcommand{\figlabel}[4][0.2]{\put(#2,#3){\parbox{#1\unitlength}{\centering #4}}} 

\usepackage{soulutf8}
\usepackage{amsmath}
\usepackage[utf8x]{inputenc}
\usepackage[margin=1in]{geometry}
\usepackage{float}
\usepackage{wrapfig}
\usepackage{adjustbox}
\usepackage[colorlinks,citecolor=red,urlcolor=blue,bookmarks=false,hypertexnames=true]{hyperref} 
\usepackage{float}
\usepackage{natbib}
\newcommand{\HM}[1]{{\color{black}{#1}}}
\newcommand{\HMR}[1]{{\color{black}{#1}}}

\begin{document}

\title{Cavitation Rheology of Model Yield Stress Fluids Based on Carbopol}

\author{Hadi Mohammadigoushki}
\email[Corresponding Author: ]{hadi.moham@eng.famu.fsu.edu}
\affiliation{Department of Chemical and Biomedical Engineering, FAMU-FSU College of Engineering, Florida State University, Tallahassee, FL, 32310, USA}
\author{Kourosh Shoele}
\affiliation{Department of Mechanical Engineering, FAMU-FSU College of Engineering, Florida State University, Tallahassee, FL, 32310, USA}

\date{\today}

\begin{abstract}

Measuring surface tension of yield stress fluids has remained a critical challenge due to limitations of the traditional tensiometry techniques. Here, we overcome those limits and successfully measure the surface tension and mechanical properties of a model yield stress fluid based on Carbopol gels via a needle-induced cavitation (NIC) technique. Our results indicate that the surface tension is approximately 70$\pm$3 mN/m, and is independent of the rheology of yield stress fluid over a wide range of yield stress values $\sigma_y = 0.5-120$ Pa. In addition, we demonstrate that a Young modulus smaller than $E<$1 kPa can be successfully measured for Carbopol gels with NIC method. Finally, we present a time-resolved flow structure around the cavity in a host of yield stress fluids, and assess the impact of fluid rheology on the detailed form of flow around the cavity. Interestingly, prior to the critical point associated with cavitation, the yield stress fluid is weakly deformed suggesting that the measured surface tension data reflect the near equilibrium values. Beyond the critical point, the yield stress fluid experiences \HMR{a strong flow} that is controlled by both the critical pressure and the non-Newtonian rheology of the yield stress fluid.

\end{abstract}


\maketitle 


\section{Introduction}
\vspace{-0.5cm}
Yield stress fluids are common in our daily life and have been the subject of intense research in the past decades\cite{frigaard2019,balmforth2014,coussot2014,malkin2017}. Prime examples include food stuff, paint, home care products, drilling fluids, oil extraction products, concrete as well as other advanced functional materials such as colloidal gels\cite{saud2021}, emulsions\cite{foudazi2015}, soft glassy materials\cite{ovarlez2010} and jammed suspensions\cite{gross2014}. In the classical description, yield stress materials behave like a solid (or deform in a finite way) below a critical stress threshold known as yield stress and flow beyond this stress threshold\cite{Bing22}. Yield stress fluids commonly interact with solid substrates\cite{nelson2020}. For example, application of yield stress lotions to skin, painting the wall, 3D printing of polymer melts on substrates. The performance of the yield stress material in these applications is controlled by its wetting and surface tension. In addition, surface tension of the yield stress fluids plays a critical role in environmentally important applications such as oil-sand pond reclamation\cite{SMALL2015,pourzahedi2021} and nuclear waste management\cite{johnson2017,kam2001}. For a simple liquid and at equilibrium the viscous stresses are negligible and provided that the gravitational forces are small, an equilibrium surface tension can be quantified\cite{israelachvili2011}. Unlike simple liquids, measuring the surface tension of a yield stress fluid has remained a critical challenge mainly because at equilibrium, the residual stresses in the yield stress fluids are no longer negligible. \par

Carbopol gels provide a model yield stress fluid for surface tension analysis and \HM{have} been the subject of several studies\cite{HU1991723,MANGLIK200155,G_raud_2014,Bouj13,C5SM00569H,Lopez18}. Carbopol gel is a transparent system, and its rheological properties can be fine tuned by variation of Carbopol concentration or solution pH\cite{gutowski2012}. This material is generally considered as a non-thixotropic yield stress fluid\cite{balmforth2014,jaworski2021}, making this system a model yield stress fluid for studies of surface tension. The earliest attempt in measuring the surface tension in Carbopol solutions goes back to the work of Hu et al.\cite{HU1991723}. These authors used a maximum bubble pressure (MBP) method, and reported a surface tension of approximately around the measured value for pure water (\HM{$\gamma$} $\approx$ 72.5 mN/m) for their solutions\cite{HU1991723}. A closer inspection of their rheological data reveals that the Carbopol based fluids used by these authors \HM {(Carbopol 934 at concentrations between 0.05-0.1 wt$\%$)} did not exhibit a yield stress rheology\cite{HU1991723}. Manglik et al.\cite{MANGLIK200155} also used the MBP method and measured the surface tension of a range of Carbopol solutions with concentrations ranging from 0 to 2000 ppm, and showed that the surface tension in this range of concentration is also close to the surface tension of water~\cite{MANGLIK200155}. However, the latter study did not report the details of Carbopol neutralization and the rheology data, which makes it unclear if those solutions were yield stress fluids\cite{MANGLIK200155}. More recently, Geraud et al.~\cite{G_raud_2014} used a capillary rise experiment to measure the surface tension of a yield stress fluid based on Carbopol gel. These authors noted a surface tension of 51 mN/m for their system\cite{G_raud_2014}. Boujlel and Coussot~\cite{Bouj13} used a Wilhelmy plate and suggested that this device generates an equilibrium surface tension around 66 mN/m at vanishingly small Capillary numbers~\cite{Bouj13}. \HMR{The Capillary number is defined as the ratio of the yield stress to the surface tension.} Jørgensen et al.~\cite{C5SM00569H} used a bridge tensiometer that involved compression and extension of the yield stress fluids between two walls. These authors showed that the surface tension data obtained from compression experiments \HM{are} smaller than those measured during extension and this deviation increases as the yield stress of the material increases~\cite{C5SM00569H}. More recently, Lopez and co-workers used a ring tensiometer and reported a surface tension value of 73.4 mN/m for their Carbopol based yield stress fluids~\cite{Lopez18}. A summary of the above literature is reported in Table~(\ref{tbl:literature}). \par 

Apparently, there is no consensus on the value of the surface tension in yield stress fluids. The complications associated with the use of solid substrates may play a significant role in this disparity. In particular, majority of prior experiments have measured surface tension by using methods that involve a contact between solid substrates and the yield stress fluid (e.g., Wilhemly plate or Bridge tensiometer). It is very well known that yield stress fluids are susceptible to wall-slip~\cite{jalaal2015,walls2003,bertola2003,divoux2011,abbasi2021}, and therefore, the fluid contact-line on such substrates may not be pinned during surface tension measurements. In particular, Geraud et al.~\cite{G_raud_2014} briefly noted that in capillaries with smooth surfaces, the yield stress fluid shows significant wall-slip. However, Jørgensen et al.~ \cite{C5SM00569H} and Boujlel $\&$ Coussot ~\cite{Bouj13} neither specified the type of surfaces used in their experiments, nor investigated the impact of wall-slip on their surface tension data.\par 
\begin{table*}[h!]

\begin{tabular}{c c c c c c}

\hline
Carbopol Type & \HM{Concentration (wt$\%$)} &$\sigma_y$ [Pa] & Technique used & $\gamma$ [mN/m] &  Reference  \\ \hline
\hline
 934 & \HM{0-0.2}& --& Maximum Bubble Pressure & 73 &\cite{MANGLIK200155} \\
934 & \HM{0.05-0.1}  &-- & Maximum Bubble Pressure & 72.5 & \cite{HU1991723} \\
ETD 2050 &\HM{0.5} & 4 & Capillary Rise & 52 &\cite{G_raud_2014} \\

ETD 2050 & \HM{0.25-2} & 0.3-38 & Bridge Tensiometer & $\approx $15-100& \cite{C5SM00569H} \\
ETD 2050 & \HM{0.25-0.5}  & 0.3-1.75 & Ascending Bubble & 59-66 &\cite{C5SM00569H} \\
 U 10 & \HM{0.1-0.5} & 9-80 & Wilhelmy Plate & 66 & \cite{Bouj13} \\

980 & \HM{0.09-0.2} & 3-28.5& LAUDA Tensiometer & 73.4 &\cite{Lopez18} \\
\hline

\end{tabular}
\caption{Summary of previous studies that have measured the surface tension for Carbopol gels. \HM{Here $\sigma_y$ and $\gamma$ denote the yield stress and the surface tension of the fluid.}}
\label{tbl:literature}
\end{table*}

Alternatively, there exists other methods for surface tension measurements that do not rely on fluid contact with a solid substrate. Examples include drop weight\cite{yildirim2005}, pendant drop\cite{stauffer1965,lecacheux2022} and \HM{maximum bubble pressure (MBP)}\cite{mysels1990,fainerman2004,simon1851} methods. Although drop weight and pendant drop methods have been well established for simple fluids, these methods are not suitable for yield stress fluids not only because the residual yield stresses affect the drop behavior compared to simple liquids, but also the basic theories do not incorporate the impact of residual yield stresses in the flow analysis for calculation of the surface tension.

\HM{The MBP method was first introduced by Simon in 1851~\cite{simon1851},} and has been widely used to measure the surface tension of a broad range of simple liquids~\cite{kuffner1961,austin1967,joos1981,fainerman2004}. In the MBP experiments a capillary tube (or a needle) is immersed in the fluid, and subsequently a gas is pumped through the needle into the liquid. \HM{As gas is injected into the needle, the gas-liquid interface forms a curved shape and its curvature is controlled by the pressure difference between the gas inside the needle and the surrounding fluid}. 
\HM{In principle, for any fluidic environment, the pressure inside the needle (or cavity) $P$ is balanced by the outside pressure and can be obtained as: 
\begin{equation}
    P = P_h + \gamma (1/r_1 + 1/r_2) + P_{out},
         \label{eq_LAP}
\end{equation}
where $P_h = \rho g z$ is the hydrostatic pressure at the needle tip, $\gamma$ is the surface tension, $r_1$ and $r_2$ \HMR{are the principal radii of curvature of the cavity} and $P_{out}$ is the pressure of the surrounding medium \HMR{resisting} against motion and growth of the cavity. Experimental observations in simple liquids have shown that as the gas is injected into the needle, the total pressure inside the needle increases and at some critical point, it goes through a maximum and suddenly drops to small values\cite{fainerman2004,mysels1990,mysels1986improvements}. This maximum pressure has been used to evaluate the surface tension of the liquids\cite{fainerman2004,mysels1990,mysels1986improvements,mysels1989some}. However, to obtain the  surface tension from the maximum pressure data, one has to account for multiple factors (i.e., curvature of the cavity, Buoyancy as well as hydrodynamic stresses in the liquid). Previous studies have shown that in the limit of small needle sizes $ R < 1 $mm, a spherical cavity forms at the tip of the needle\cite{fainerman1998maximum}. In particular, at the maximum pressure ($P_c$), the cavity forms a hemispherical shape with a radius equal to that of the needle radius, and therefore, Eq.~\ref{eq_LAP} can be expressed as: $P_{c} - P_h = \frac{2\gamma}{R} + P_{out}$. Any gas injection beyond this point causes the cavity to become unstable, and grow rapidly before detaching from the needle. For needles with $R >1 mm$, the cavity becomes non-spherical\cite{sugden1924determination}. It has been shown that the impact of cavity non-sphericity on the surface tension could be accounted by introducing a correction factor $f$ into the above equation such that: 
\begin{equation}
P_c - P_h = f \frac{2\gamma}{R}+ P_{out}.
\label{laplace}
\end{equation}

The correction factor can be obtained as\cite{fainerman1998maximum}: \HMR{$f = {\sum_{i=0}^{5}} a_{i}{(\frac{R}{a})}^{i},$} where capillary length $a = \sqrt{\frac{2\gamma}{\Delta \rho g}}$. 
%
Note $a_i$ values are tabulated in the literature~\cite{fainerman1998maximum,bendure1971dynamic}. For viscous liquids $P_{out}$ is associated with the viscous stresses. Previous studies have investigated the impact of hydrodynamic stresses on surface tension data of a range of viscous fluids~\cite{fainerman2004correction}. Fainerman and co-workers showed that the viscous stresses can increase the surface tension by up to 3 mN/m over two orders of magnitude variation in the viscosity of a solution with an equilibrium surface tension of 70 mN/m, thereby, suggesting a small influence of the viscous stresses on equilibrium surface tension of viscous liquids~\cite{fainerman2004correction}.}

\section{Cavitation Rheology}

Yield stress fluids are different from simple liquids in that below the yield stress threshold, they behave like a solid and barely deform. As a result, the gas pressure inside the capillary tube must overcome the residual stresses in the solid before they can plastically deform the surrounding material. Over the course of last decade, Crosby and co-workers have developed a needle-induced cavitation (NIC) rheology technique based on the MBP method for a wide range of synthetic hydrogels, rubbers and block co-polymers~\cite{zimberlin2007,kundu2009,zimberlin2010,barney2020,cui2011,delbos2012,hutchens2016,barney2019}. \HM{It has been shown that for elastic materials, and in the limit of small needle sizes, similar to simple liquids, the maximum pressure occurs at the point where cavity forms a hemisphere at the tip of the needle and the critical pressure $P_c$ is related to surface tension and Young modulus $E$ of the gel as~\cite{zimberlin2007,gent1969}:
\begin{equation}
     P_c = 5E/6 + 2\gamma/R,
     \label{eq_NIC}
\end{equation}
where, the elasticity of the surrounding material acts as an additional pressure that cavity must overcome with $P_{out} = 5E/6$. Therefore, the maximum pressure measurement in the MBP method not only allows one to estimate the surface tension of the gel, but also its elastic modulus, hence the term cavitation rheology was used.} \HM{Note that in Eq.~(\ref{eq_NIC}) the surrounding material behaves as a neo-Hookean solid with no plastic deformation. As a result, the impact of local yielding of the materials on surface tension data is neglected. The latter assumption may hold for stiff materials such as synthetic hydrogels, rubbers and block co-polymers that are stiff and exhibit strong elasticities. The yield stress fluids based on Carbopol gels are expected to be much softer than synthetic hydrogels, acrylic triblock gels or copolymers used in previous studies~\cite{zimberlin2007,barney2019}. For yield stress materials based on Carbopol, the $P_{out}$ resistance may be related to both the elastic response of the yield stress fluid before reaching plasticity as well as the plastic response of the medium after it passes the yield limit in the vicinity of a growing bubble. Consequently, Eq.~(\ref{eq_NIC}) should be modified for soft yield stress fluids prepared from Carbopol solutions. } 

\HM{To the best of our knowledge, the theoretical analysis of the cavitation phenomenon in soft yield stress fluids has not been considered before. Additionally, there are currently no studies that employ the NIC technique to measure the surface tension of the yield stress fluids based on Carbopol gels.} Although previous studies have used the NIC method to measure the Young modulus of a wide range of stiff materials with a modulus in the range of 1 kPa $ < E <$ 60 kPa\cite{barney2019}, the yield stress fluids based on Carbopol gels are expected to be much softer than stiff gels and co-polymers~\cite{barney2019}. \HM{Hence, it is still unclear if the NIC technique is sensitive enough to measure a Young modulus in the range of E $<$ 1 kPa that is relevant for Carbopol gels.} \HM{The main goal of the first part of this paper is to develop a theoretical framework for cavitation in soft yield stress fluids and its subsequent application in experiments to evaluate the surface tension and the mechanical properties of the yield stress fluids based on Carbopol gels. }\par 

On a relevant note and from a biological perspective, a very important health issue, traumatic brain injury (TBI), has been associated with the cavitation of a bubble in biological tissues that exhibit yield stress properties~\cite{barney2020}. The leading hypothesis is that the cavitation introduces a strong flow, which consequently deforms and damages the surrounding tissue, and that causes TBI\cite{canchi2017}. To test this hypothesis, one must first evaluate the flow field generated by a cavitation process in a yield stress material. A direct and in situ measurement of the flow profile around a cavity in yield stress materials do not exist. Imaging the detailed form of flow structure around a cavity in a model yield stress fluid provides important insights that will significantly advance our understanding of the origin of the TBI. The main goal of the second part of this paper is to provide the first temporally-resolved form of flow structure around a cavity in model yield stress fluid based on Carbopol gels.

\section{Materials and Methods}


Yield stress is observed in a host of materials. The most popular yield stress fluid is a polymeric gel based on an aqueous solution of Carbopol\cite{balmforth2014}. Different models of Carbopol (940, 980, Ultrez-10 and etc) are commercially available. In this study we use Carbopol 940 and Ultrez-10 which are known as non-thixotropic model yield stress fluids\HM{~\cite{balmforth2014}}. The concentration of the Carbopol is varied from 0.02 wt$\%$ to 0.5 wt$\%$. Yield stress fluids are made by gently mixing Carbopol with de-ionized water and neutralized by adding 1.5 mass fraction of triethanolamine to the Carbopol solution. In addition to Carbopol based fluids, we use a Newtonian fluid \HM{(corn syrup from Golden Barrel)} for the purpose of comparison with the yield stress fluids \HM{(see rheological properties in Table~S(1) of the supplementary materials)}. \par

Yield stress fluids were characterized using a commercial rheometer (TA Discovery HR 10) and a standard concentric cylinders geometry with R$_i$ =14.01 mm and R$_o$ = 15.185 mm, where R$_i$ and R$_o$ are the radii of the inner and outer cylinders. Because wall-slip is significant and can affect the rheological characterization of the yield stress fluids, we have roughened the concentric cylinders geometry using a sand blasting protocol. As in our previous studies~\cite{Rassolov2020,Rassolov2022,nazari2023helical}, two types of measurements are performed: Small Amplitude Oscillatory Shear (SAOS) was used to obtain the linear viscoelastic responses. In particular, the storage and loss moduli are measured as a function of angular frequency. In addition, the flow curves are measured using a steady applied shear experiment.\par

\begin{figure}[h]
\centering
\vspace{-5cm}
 \setlength{\unitlength}{0.9\textwidth}
\begin{picture}(1,1)
    \put(0,0){\includegraphics[width=0.8\textwidth]{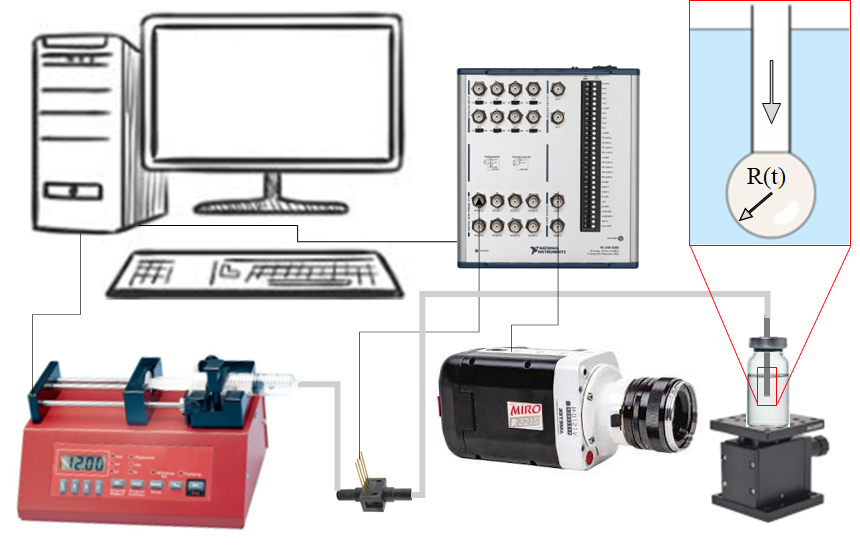}}
     \figlabel[0.4]{0.595}{0.58}{Gas}
    \figlabel[0.2]{0.05}{0.0}{Syringe Pump}
    \figlabel[0.35]{0.25}{0.03}{Pressure Sensor}
     \figlabel[0.25]{0.155}{0.48}{Computer \& Control Unit}
     \figlabel[0.4]{0.35}{0.53}{Data Acquisition Unit}
    \figlabel[0.4]{0.42}{0.06}{High Speed Camera}
     
    \end{picture}
	\caption{\small A schematic of the NIC device along with the necessary parts used in this paper. }\label{NIC}
\end{figure}

To measure the surface tension and the Young modulus of the yield stress fluids, we use an in-house custom-made NIC method. Fig.~\ref{NIC} shows a schematic of the custom made NIC technique, which consists of a programmable syringe pump (model NE-1000 from New Era), blunt needle (from McMaster-Carr), tubing, wiring, differential pressure sensor (model PX-26 series from OMEGA), high-speed camera (Phantom miro M310), data acquisition device; DAQ (from National Instruments), and a computer. The LabVIEW software and the DAQ allow the differential pressure sensor, programmable syringe pump, and high-speed camera to work in a synchronized manner. Using this setup, we are able to record, and capture the temporal evolution of cavity growth and the associated pressure changes. The inner radius of the needle (or the capillary tube) used in these experiments is varied from \HM{$R$ =  76 $\mu$m - 850 $\mu$m}, which allows us to access a wide range of critical pressures. In each experiment, the needle is gently placed in the yield stress fluid and the air is injected in the capillary tube at a constant rate of \HM{$Q = 0.3~\mu$L/hr}. \HMR{Fluids are placed in a cubical container with flat side walls to minimize optical  distortions.} In addition to NIC method, we use a pendant drop technique to measure the surface tension of primarily non-yield stress fluids (see more details about this technique in the supplementary materials). The surface tension data obtained from the pendant drop experiments (for simple liquids) are compared with the results obtained by the NIC method to ensure the accuracy of the NIC method.  \par 

We have also performed particle image velocimetry (PIV) to temporally resolve the flow field around the cavity in Carbopol gels. For PIV analysis, we generate a sheet of laser light (with a wavelength of 532 nm) that passes through the cavity. The fluids are seeded by glass microspheres (model 110P8 from Potters with a mean diameter $\approx 8~\mu$m). The small amount of these seeding particles (40 ppm by mass) does not affect the rheological properties of the fluids. \par

\vspace{-1cm}

\section{Results and Discussion}

\subsection{Bulk rheology} 
\begin{figure}[h]
\centering
	\includegraphics[width=1\textwidth]{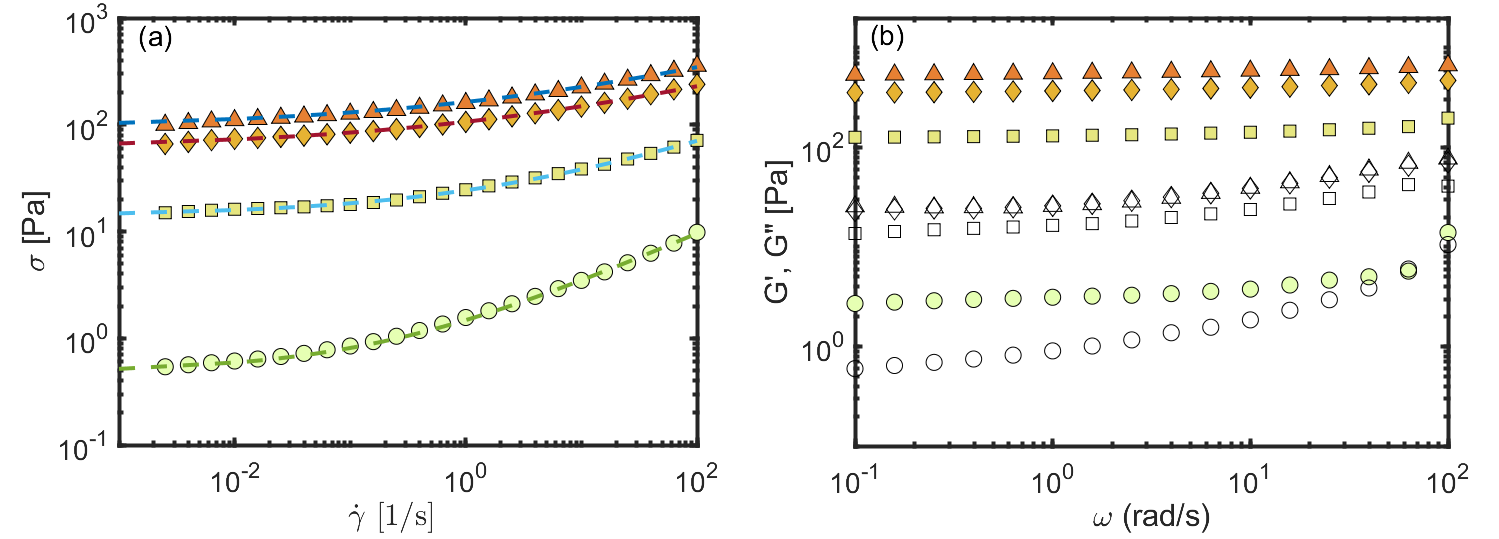}
	\caption{\small (a) steady shear stress as a function of applied shear rate for yield stress fluids based on Carbopol Ultrez-10. (b) Storage (filled symbols) and loss moduli (open symbols) as a function of angular frequency. Different symbols correspond to various Carbopol concentration as 0.06 wt$\%$ ($\circ$), 0.1 wt$\%$ ($\square$), 0.3 wt$\%$ ($\diamond$) and 0.5 wt$\%$ ($\triangle$).}\label{rheology}
\end{figure}

Fig.~\ref{rheology}(a) shows the flow curves of sample representative yield stress fluids based on Ultrez-10 along with the best fit to the Herschel-Bulkley fluid model (dashed curves). The Herschel-Bulkley model is defined as: $\sigma = \sigma_y + K{\dot{\gamma}}^{n}$, where $\sigma$, $\sigma_y$, $K$, $\dot{\gamma}$ and $n$ denote the shear stress, yield stress, consistency factor, rate of deformation and the shear-thinning index, respectively. As expected, the increase in the Carbopol concentration gives rise to a stronger yield stress fluid. In addition, Fig.~\ref{rheology}(b) shows the measured storage and loss modulii as a function of angular frequency for sample yield stress fluids based on Ultrez-10. As the Carbopol concentration increases both the storage and loss moduli increase, which is again expected \HM{and consistent with the reported values in the literature\cite{gutowski2012}}. Moreover, the storage modulus approaches an asymptotic value (called shear modulus hereafter) as the angular frequency decreases. Similar rheological properties have been reported for the yield stress fluids based on CBP-940 and a summary of the rheological properties of the Carbopol solutions are included in the table~(S1) and Fig.~S1 of the supplementary materials. \HMR{In addition, the shear viscosity of these fluids is not sensitive to ramp-up or ramp-down in shear rate (see Fig.~S2 in the supplementary materials), thereby confirming that these yield stress fluids are not thixotropic.} \par 

\begin{figure}[h]
\centering
	\includegraphics[width=1\textwidth]{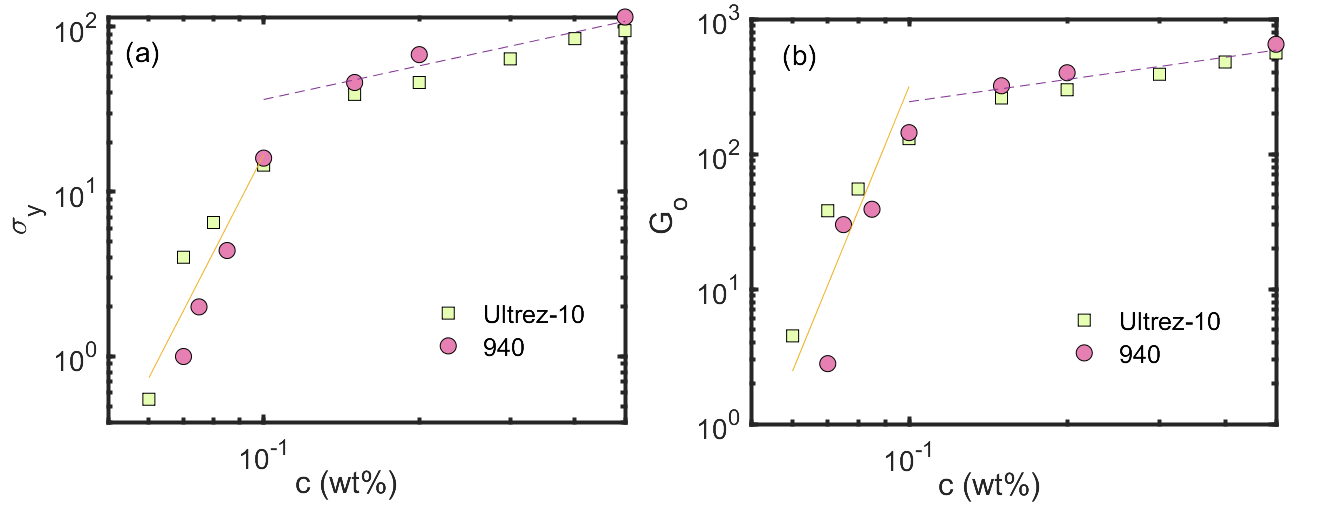}
	\caption{\small (a) Yield stress and (b) shear (elastic) modulus as a function of Carbopol concentration for the two yield stress fluid systems. The continuous and dashed lines are guide to the eye.}\label{rheology2}
\end{figure}

Fig.~\ref{rheology2} shows a summary of the yield stress and the shear modulus of these yield stress fluids as a function of Carbopol concentration. Interestingly, at low Carbopol concentrations, the yield stress and the shear modulus increase rather sharply as the Carbopol concentration increases. However, beyond a critical concentration (between 0.1-0.15 wt$\%$), the increase in yield stress and the shear modulus becomes more gradual. This trend can be rationalized as follows. Carbopol is made up of high molecular weight polymers that swell upon mixing and neutralization in aqueous solutions\cite{Oel22}. At low Carbopol concentrations, the swollen polymer particles form a percolated structure, which at higher concentrations, the polymer swelling increases, thereby, decreasing the distance between polymer particles in the solution. The decrease in particle spacing increases the yield stress and the shear modulus. Beyond a critical point, where particles create a jammed structure, further increase of the Carbopol content does not dramatically affect the jammed nature of the solution and therefore, the yield stress and the elastic modulus increase more gradually.   
\subsection{Needle induced cavitation}

\begin{figure}[h]
\centering
	\includegraphics[width=1\textwidth]{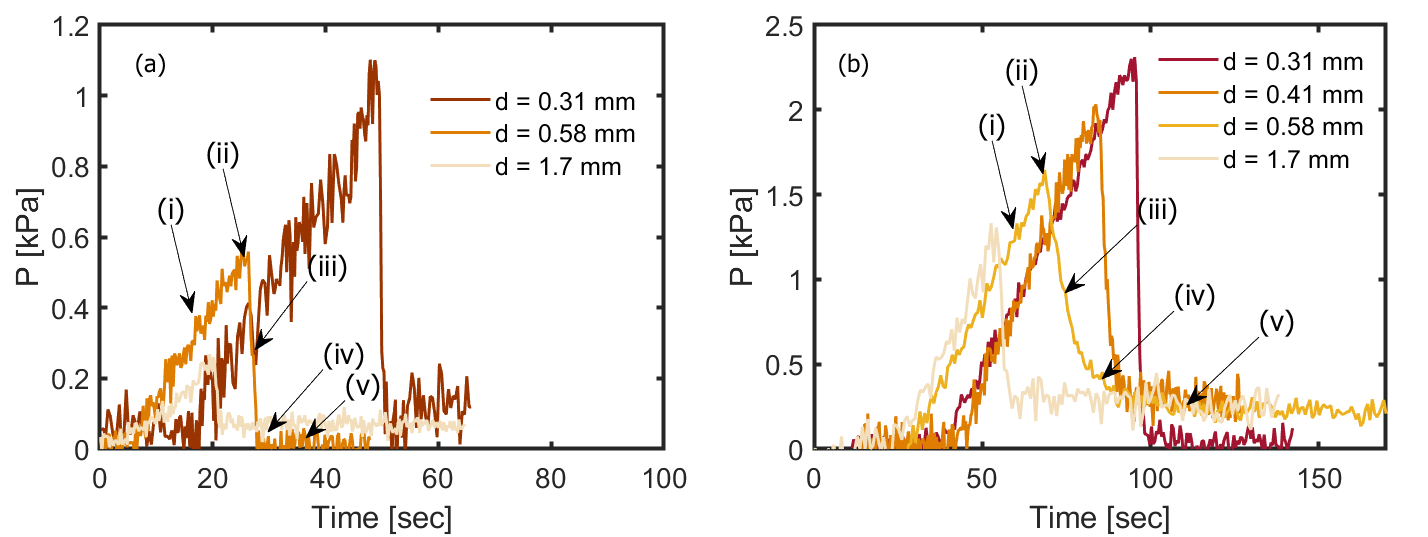}
		\includegraphics[width=1\textwidth]{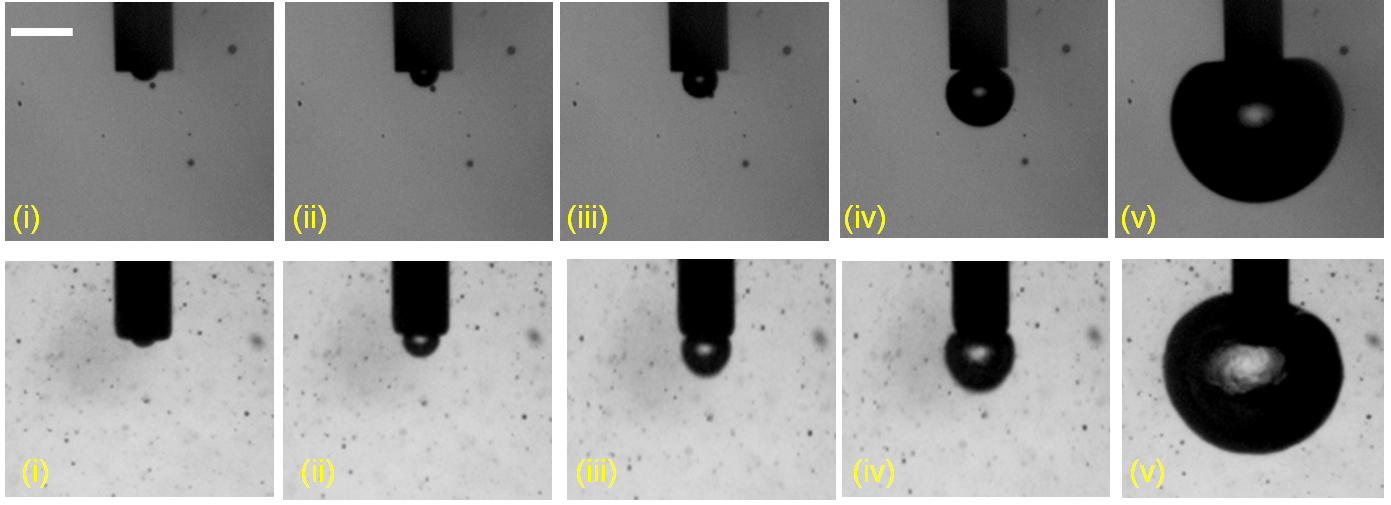}
	\caption{\small Temporal evolution of the pressure in (a) corn syrup and (b) 0.5wt$\%$ yield stress fluid based on Ultrez-10. The transient pressure is shown for various needle tube diameters $d$. The lower snapshots show the simultaneous temporal evolution of cavity in corn syrup (top row) and in 0.5wt$\%$ yield stress fluid based on Ultrez-10 (lower row). \HM{Note that (i-v) in snapshots refer to the time instances at which pressure is measured before or after the critical point in part (a) and part (b). \HMR{The scale bar in snapshot is 0.8 mm. }}}\label{P_time}
\end{figure}

\HM{As noted above, the maximum pressure inside the capillary tube in soft yield stress fluids may be related to surface tension, the yield stress and the elastic modulus of the material. Therefore, in the first step, we developed a mechanical analysis of the cavitation in yield stress fluids. In the limit of small capillary size ($R< 1mm$ or for a nearly spherical bubble), in addition to the surface tension pressure $2\gamma/R$, the effective pressure acting on the yield materials surrounding the cavity $P_{out} = P_{out}(\sigma_y,E)$. In the near field, adjacent to the bubble, the material may yield and far away from the bubble, the yield stress material behaves as an elastic solid. Considering the yield stress material as a linear elastic perfectly plastic material, the total pressure acting on the cavity \HMR{$P_{out}$ = $P_{out}|_{1}$ + $P_{out}|_{2}$. Here $P_{out}|_{1}$ and $P_{out}|_{2}$ are pressures associated with plastically deformed response of the surrounding medium near cavity and the confinement induced by the elastic response of the yield stress material, respectively. Based on our analysis, the total pressure contribution from the surrounding yield stress material on the cavity can be written as (see appendix for more details on the derivation):
\begin{equation}
P_{out}\,=\frac{2\sigma_y}{3} \left\{1+\ln\left(\frac{2E}{3\sigma_y}\right)\right\}+\frac{2\pi^2}{27} E.
\label{eq_pcritical31}
\end{equation}
Note the latter term in total pressure ($P_{out}|_2 = \frac{2\pi^2}{27} E$) is analogous to the Eq.~\ref{eq_NIC} derived for elastic networks\cite{zimberlin2007,gent1969}.
Therefore, in the limit of small capillary tube diameters, the maximum pressure inside the growing bubble in a soft yield stress material can be given as: 
\begin{equation}
P_{c}\,= \frac{2\gamma}{R} + \frac{2\sigma_y}{3} \left\{1+\ln\left(\frac{2E}{3\sigma_y}\right)+ \left(\frac{\pi}{3}\right)^2\frac{E}{\sigma_y}\right\}.
\label{eq_pcritical3_t}
\end{equation}
}

The above Eq.~\ref{eq_pcritical3_t} will be used to assess the surface tension as well as the elastic modulus of the yield stress fluids based on Carbopol.} \HM {Subsequent to the above theoretical analysis,} we performed needle-induced cavitation experiments in a Newtonian corn syrup as well as yield stress fluids. Fig.~\ref{P_time} shows the temporal evolution of the pressure in the Newtonian solution based on corn syrup (Fig.~\ref{P_time}(a)) and a sample yield stress fluid (0.5wt$\%$ Ultrez-10; Fig.~\ref{P_time}(b)). \HM{In these experiments, the volume of the air left in the syringe at any point in time can be given as: $V(t) = V_0 - Qt + V_b$, where $V_0$, $Q$ and $V_b$ are the initial air volume (which is 10mL in this work), flow rate by which the air is pumped and the volume of the cavity. If we assume that air is an ideal gas and experiments are performed in isothermal conditions, $PV$ = $P_0V_0$ and therefore, we will have: 
\begin{equation}
    {\Delta P}/{P_0} ={(P- P_0)}/{P_0} = {1}/({1-\frac{Q}{V_0}t + \frac{V_b}{V_0}})\,-\,1
\end{equation}
In the limit of $V_b << V_0$, and $Qt<V_0$ (which are satisfied in our experiments) pressure should increase linearly with time regardless of the type of fluid. Indeed, our results of Fig.~\ref{P_time}(a,b) show that for both Newtonian and the yield stress fluids, the pressure increases linearly up to a critical value before it drops quickly.} Additionally, the critical pressure increases as the needle diameter decreases, which is consistent with the predictions of Eq.~(\ref{laplace}) for the Newtonian fluid and Eq.~(\ref{eq_pcritical3_t}) for the yield stress fluid.  

\HM{The corresponding temporal evolution of the cavity size are shown for the corn syrup (top row in Fig.~\ref{P_time}) and the yield stress fluid (bottom row in Fig.~\ref{P_time}). As the pressure increases towards the critical point, the cavity gradually protrudes out of the needle tip and attains a smaller radius of curvature. At the critical point, the cavity forms a hemisphere with a diameter that is approximately equal to the diameter of the needle for both Newtonian and the yield stress fluids. This is unsurprising because the correction factor $f$ in these experiments is very close to unity ($f = 0.9968$). Beyond the critical pressure, the explosive growth of the cavity radius gives rise to the pressure drop. The experiments of Fig.~\ref{P_time} are performed in all solutions and the measured critical pressure values along with the best fit to Eq.~\ref{eq_pcritical3_t} could be used to assess the surface tension and the Young modulus of the yield stress fluids.} Prior to assessment of the yield stress fluid properties, and in order to ensure the accuracy and precision of our NIC device, we first report the results for a broad range of non-yield stress fluids.\par 

\begin{figure}[h]
\centering
	\includegraphics[width=0.5\textwidth]{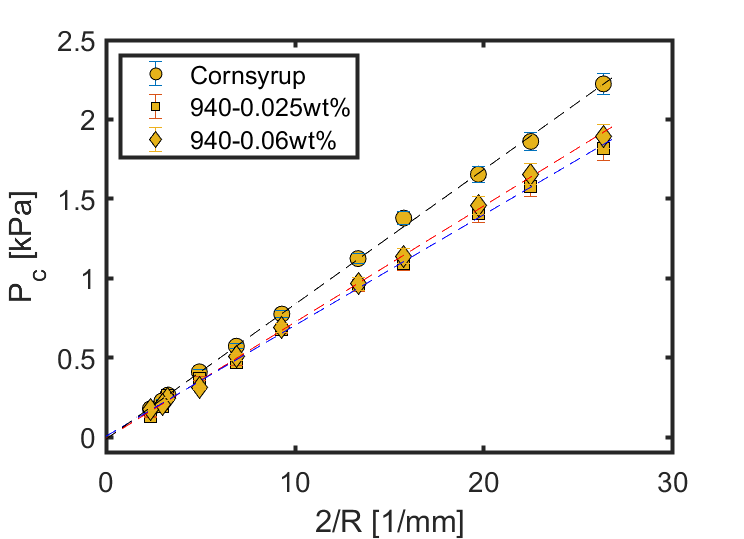}
	\caption{\small Critical pressure as a function of capillary tube size for non-yield stress fluids. The dashed lines indicate the best linear fit to the experimental data. }\label{Pd_callibration}
\end{figure}

Fig.~\ref{Pd_callibration} shows the critical pressure as a function of needle diameter for corn syrup along with Carbopol based solutions that are in the dilute regime and show non-yield stress rheology (see Table~(S1) in the supplementary materials for rheological properties). \HM{Subsequently, the results were fitted to Eq.~(\ref{laplace}). In experiments performed in this study, the maximum needle radius used is $0.85$mm and therefore, the maximum correction factor due to non-sphericity of the cavity is about $f = 0.97$. Additionally, unlike previous MBP studies that have mainly used one needle size to assess the surface tension of liquids, we fit the critical pressure data to a wide range of needle sizes and that reduces the error associated with using only one measurement point.} The resulting intercepts and slopes are summarized in Table~(\ref{table:2}). \par 
\HM{The first notable observation is that $P_{out}$ contribution from viscous stresses is very small ($\leq$ 8 Pa) both for the corn syrup as well as other non-yield stress fluids based on Carbopol. Note that the maximum $P_{out}$ measured in our experiments is smaller than the pressure associated with the surface tension of the liquid by orders of magnitude. For example, for the largest needle used, $R = 0.85$mm, the surface tension pressure is $\approx$ 0.17 kPa while $P_{out} = 0.0004-0.008$ kPa indicating a negligible impact of the viscous stresses on surface tension measurements. In addition, although some of these viscous liquids (cf. Carbopol 940 0.025 wt $\%$ and 0.05 wt $\%$ in Table~S(1) of the supplementary information) have significantly different shear viscosity, the resulting $P_{out}$ is still negligible and similar in magnitude. We conclude from these results that the impact of viscous stresses on surface tension data is negligible. This conclusion is consistent with previous findings on viscous fluids~\cite{fainerman2004correction}. } \par 
Secondly, the slope of each graph, which represents the estimated surface tension of these fluids are listed in Table~\ref{table:2}. In particular, for corn syrup, the measured surface tension is consistent with the value reported in the literature~\cite{llewellin2002,borhan1999}. In addition, for dilute Carbopol solutions the surface tension is very close to \HMR{that measured} for pure water, independent of the type and concentration of the Carbopol. The latter data are consistent with measurements of Manglik et al~\cite{MANGLIK200155} and Hu et al.~\cite{HU1991723} who have reported a surface tension of $\approx $ 72.5-73 mN/m for dilute Carbpol solutions based on Carbopol 934. Finally, included in the Table~\ref{table:2} are the measured surface tension data of these fluids by a pendant drop method (see details in \HMR{Fig.~S3} of the supplementary materials). The resulting surface tension \HMR{values} obtained by the pendant drop method are consistent with the NIC results, thereby, confirming the accuracy of our NIC method for evaluating the surface tension of simple liquids. \par

\begin{table*}[h!]
\begin{tabular}{c c c c c c}

\hline\hline
 Fluid & [wt$\%$] &$P_{out}$ [kPa] & $\gamma$ [mN/m] & $\gamma^*$ [mN/m] & $\gamma^{**}$ [mN/m] \\ [0.5ex] 
 \hline
 corn syrup        &  & \HM{-0.008}  & \HM{83.2} & 81~\cite{borhan1999}  & 80.5$\pm$2     \\
\multirow{2}{3cm}{\centering 940}& 0.025 & \HM{-0.006}   & \HM{70}        & $--$& 71.0$\pm$2  \\  
 &0.05 & \HM{-0.005}    & \HM{72}       & $--$ & 72.2$\pm$1   \\ 
 \multirow{2}{3cm}{\centering Ultrez$-$10}& 0.04 & \HM{-0.004}   & \HM{71.1}        & $--$ & 70.9$\pm$2 \\  
 &0.05 & \HM{0.004}    & \HM{72.1}       & $--$&  72.2$\pm$3   \\ [1ex] 
 \hline
\end{tabular}
\caption{A summary of the surface tension and fitting to the Newtonian as well as dilute Carbpol based solutions. $^{*}$ denotes the surface tension data reported in the literature and $^{**}$ are the surface tension values obtained by a pendant drop method.  }
\label{table:2}
\end{table*}
\begin{figure}[H]
\centering
	\includegraphics[width=1\textwidth]{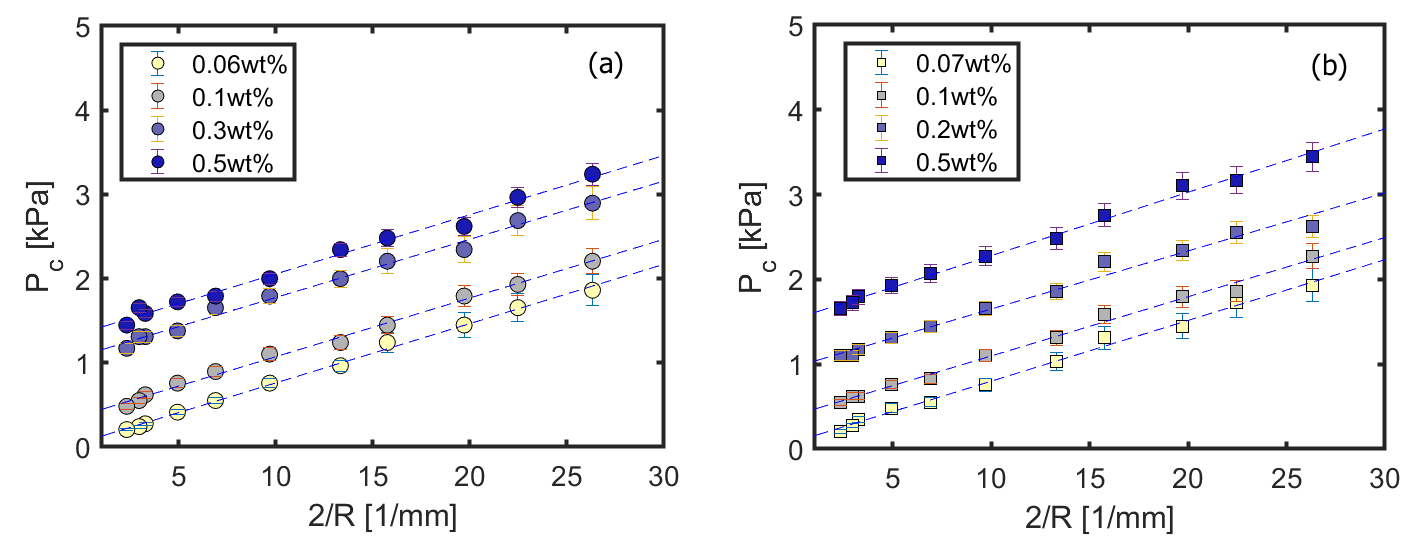}
	\caption{\small Critical pressure as a function of inverse capillary size for yield stress fluids based on (a) Ultrez-10 and (b) 940. \HMR{The dashed line is the best fit of Eq.~\ref{eq_pcritical3_t}} to the experimental data. }\label{Pd}
\end{figure}
Subsequent to the above experiments, the needle induced cavitation experiments are performed on all of the yield stress fluids and the critical pressure for a wide range of needle sizes is measured. Fig.~\ref{Pd} shows the critical pressure as a function of capillary radius for the two yield stress systems of this paper. Included in these figures is also \HMR{the best fit of Eq.~\ref{eq_pcritical3_t}} to the experimental data. First, the slope of the data in all experiments is similar hinting at a similar surface tension data for all of these systems over a broad range of Carbopol concentrations. Additionally, the intercept of the fitted line to these experimental data is no longer negligible and increases as the Carbopol concentration increases. \HM{Finally, the measured critical pressure data are independent of the imposed flow rate ($Q = 0.1-10~\mu$L/hr; see \HMR{Fig.~S4} in the supplementary materials) }and the needle insertion procedure (whether needle is gently inserted into the sample or retracted following a protocol suggested in~\cite{barney2019}). \par

\begin{figure}[H]
\centering
	\includegraphics[width=1\textwidth]{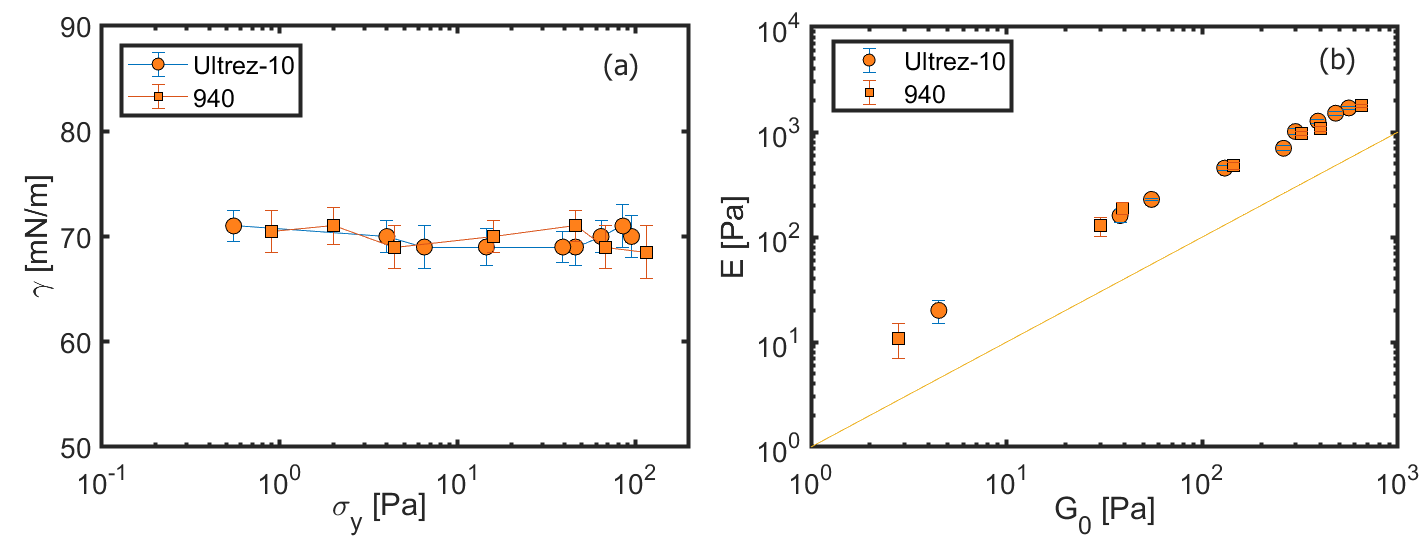}
	
	\caption{\small (a) Surface tension as a function of yield stress for yield stress fluids of this study. (b) Young modulus as a function of shear modulus for the yield stress fluids. The line in part (b) is the equivalent point. }\label{Surface}
\end{figure}

The resulting surface tension data for the two yield stress fluids are shown in Fig.~\ref{Surface}(a). Interestingly, the surface tension does not change substantially as a function of yield stress. The average surface tension data amongst all these yield stress fluids is approximately 70$\pm$3 mN/m. These results are consistent with the assumption made by Lopez et al~\cite{Lopez18} that the surface tension of the yield stress fluids does not change as the yield stress increases. Additionally, Boujlel and Coussot~\cite{Bouj13} reported the equilibrium surface tension of the yield stress fluids to be approximately 10$\%$ less than the surface tension of the pure water, which are close to our measured values as well. The difference between the data provided by Boujlel and Coussot~\cite{Bouj13} and our measured values is presumably due to potential role of wall-slip in experiments of Boujlel and Coussot~\cite{Bouj13}.  \par

In addition to surface tension, we have assessed the Young modulus, $E$ of the yield stress fluids by fitting the intercepts to the Eq.~\ref{eq_pcritical3_t}. Fig.~\ref{Surface}(b) shows the Young modulus as a function of shear modulus for all of the yield stress fluids considered in this study. For both systems, the Young modulus increases linearly with the shear modulus. \par

\subsection{Flow visualization around cavity} 

\begin{figure}[H]
\centering
	\includegraphics[width=0.9\textwidth]{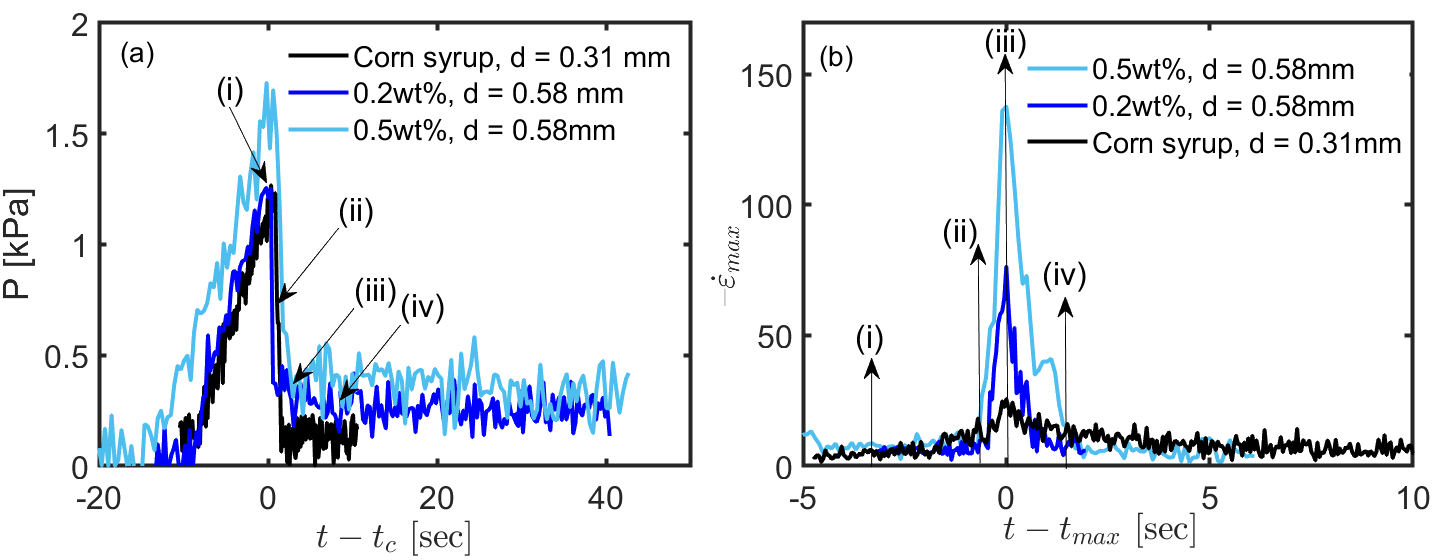}
	\includegraphics[width=0.85\textwidth]{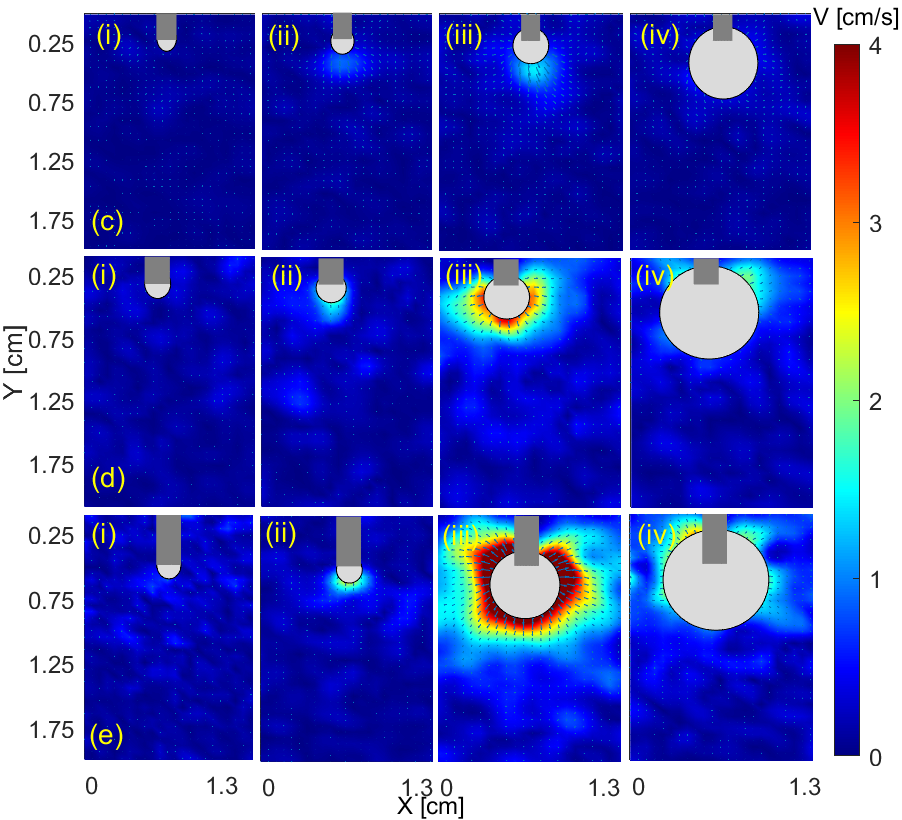}

	\caption{\small (a) Temporal evolution of pressure in NIC experiments measured for various fluids and needle sizes. Here $t_{c}$ refers to the time that pressure reaches the critical value. (b) The temporal evolution of the maximum strain rate around the cavity for various fluids. Here $t_{max}$ refers to the time that strain rate reaches the maximum value. The maximum strain rates that fluids experience over the course of cavitation are 18 [1/s] for corn syrup, 75 [1/s] for 0.2 wt$\%$ and 115 [1/s] for 0.5 wt$\%$ Carbopol gels. The temporal evolution of the velocity profiles are shown at different instances during cavity formation and growth. Each of velocity profiles corresponds to corn syrup (c), 0.2wt$\%$ Ultrez-10 (d) and 0.5wt$\%$ Ultrez-10 (e). }\label{PIV}
\end{figure}

Fig.~\ref{P_time} indicates that the cavitation experiments comprise of two steps. In the first stage, as the pressure increases linearly with time, a cavity forms at the tip of the needle and gradually decreases its radius of curvature. The second step is associated with the post critical pressure and consists of an instability that is characterized by sudden expansion of the cavity. \HM{Fig.~\ref{PIV}(a) shows the temporal evolution of the pressure for the Newtonian corn syrup along with two yield stress fluids. The time-resolved velocity profiles for these three sample experiments are shown in Fig.~\ref{PIV}(c-e) at different points along the pressure curves of the Fig.~\ref{PIV}(a). At times where the pressure values are below the critical pressure and even at the critical pressure, where the radius of the cavity is the same as the needle inner radius, the flow field around the cavity is weak (see (i) Fig.~\ref{PIV}(c-e)). Therefore, these fluids, regardless of their rheology, do not experience a strong flow at the critical point. The latter observation gives further credence to the aforementioned assumption that at the critical point, the viscous stresses are weak and their impact on surface tension is negligible. }\par 

However, post-instability (i.e., $t-t_{c} > 0$) sudden expansion of cavity introduces a strong flow to the surrounding fluid shortly after the critical point (see the velocity field in (ii) Fig.~\ref{PIV}(c-e)). Subsequently, at some point during rapid pressure drop, the flow reaches its maximum strength (see (iii) in the velocity field of Fig.~\ref{PIV}(c-e)). Eventually, the flow field subsides to equilibrium at longer times (see (iv) in Fig.~\ref{PIV}(c-e)). \HM{The above trend is observed over the range of the needle sizes used in this study both for the viscous and the yield stress fluids.} The most dominant component of the fluid flow around a cavity is the compression. Therefore, to characterize the strength of the compression around the cavity, we use the maximum extension rate around the cavity as a function of time. \HMR{Therefore, the volumetric extension rate in spherical coordinate, at each point in time, is defined as: 
\begin{equation}
    \dot{\varepsilon} = \mathbf{\nabla} \cdot \mathbf{U}=\frac{\partial u_{r}}{\partial r} + \frac{1}{r}\frac{\partial u_{\theta}}{\partial \theta} + \frac{u_r}{r}+ \frac{1}{r \sin(\theta)}\frac{\partial u_{\phi}}{\partial\phi} + \frac{u_r + u_{\theta} \cot(\theta)}{r}.
\end{equation}}

\HMR{Here $\mathbf{U}$ is the fluid velocity and $r$ is the instantaneous distance from the center of the bubble. Although flow around the cavity is three dimensional, our PIV method allows us to map the 2D velocity profiles around the cavity. Additionally, because of symmetry of the flow around the cavity, $u_{\theta} = 0$ and $u_{\phi} = 0 $. Hence, $\dot{\varepsilon} = \frac{\partial u_{r}}{\partial r} + \frac{2u_r}{r}.$ } The extension rate is obtained by first fitting a smooth function to the experimentally measured velocity profiles and then, differentiating from the smooth velocity function. At each point in time, the fluid around the cavity experiences a maximum extension rate, which typically occurs close to the cavity boundaries. We assess the flow strength in each of the above experiments by comparing the maximum extension rate that each fluid experiences during cavitation process. Fig.~\ref{PIV}(b) shows the temporal evolution of the maximum extension rate that each fluid experiences around the cavity.\par 

Interestingly, a careful comparison between the flow field of the Newtonian and the yield stress fluids of Fig.~\ref{PIV}(c-e), reveals a significant difference between these cases. Despite the critical pressure being the same for the Newtonian fluid and \HMR{0.2wt$\%$ Carbopol solution}, the velocity fields and the maximum strain rate in the yield stress fluid are stronger than the Newtonian counterpart. This difference is presumably connected to the non-Newtonian rheology of the yield stress fluid. On the other hand, as the critical pressure for the onset of cavitation increases, the flow strength (velocity field and/or the characteristic maximum strain rate) before and at the onset of cavitation (e.g., (i) in Fig.~\ref{PIV}(e)) is very negligible and similar to those shown for other systems in Fig.~\ref{PIV}(c-d). However, for post-instability, the flow field is much stronger in Fig.~\ref{PIV}(e) than those shown for other systems in Fig.~\ref{PIV}(c,d). Taken together, these results suggest that in addition to the critical pressure, the non-Newtonian rheology of the yield stress fluids control the detailed form of flow structure around a cavity.\par

\subsection{Dimensionless analysis}

\HM{Finally, we come back to \HMR{analyzing} the stresses involved in cavitation experiments. In principle, several forces can be involved in cavitation experiments; inertia, viscous, gravitational, elastic, yield stress and the surface tension. To assess the importance of these stresses, we use a range of dimensionless numbers. Note that the surface tension values are evaluated at the critical point, where pressure shows a maximum. Hence, we will assess the dimensionless numbers at the critical point.} \par

Prior and at the critical point, the deformation of the fluid, although small, is controlled by the gas injection rate. As noted before, the injection rate used in these experiments is very small and fixed for all needle sizes to $Q = 0.3~\mu$L/hr. We can estimate a characteristic deformation rate (at the walls of the capillary tube) associated with this flow rate as $\dot{\gamma} \approx {4Q}/{\pi R^3} \approx 10^{-4} - 10^{-2}$\HMR{ [1/s]}. The importance of inertia can be assessed using a Reynolds number defined as: $Re = \rho \dot{\gamma} d^2/\eta(\dot{\gamma})$. Here, $\rho$, $\dot{\gamma}$, $d$ and $\eta$ represent the density of the fluid, characteristic deformation rate, the diameter of the capillary tube and the shear dependent viscosity of the surrounding fluid. Although inertia depends on the choice of the needle and fluid, the maximum Reynolds number for all experiments is very small ($Re \approx 10^{-6}$), rendering the effects of inertia negligible. \par 
\HM{The impact of gravitational forces can be estimated by a Bond number defined as $Bo = {\Delta \rho g a^2}/{\gamma}$. At the critical point, the cavity forms a hemisphere with a radius equal to the needle size. Therefore, at the critical point, the Bond number is controlled by the needle size and varies as $Bo \approx 3\times 10^{-3} - 0.1$ for all needle sizes used in this paper. Therefore, the effect of gravity on surface tension data is negligible. The latter finding is consistent with nearly spherical bubble shapes observed in all of our experiments.} \HM{To assess the impact of viscous stresses, we first start with cavitation experiments on Newtonian and the non-yield stress fluids that suggest the viscous stresses are negligible compared to the surface tension effects. In experiments with the yield stress fluids, the viscous stresses, if strong enough, may plastically deform the material near the cavity. To assess the relative importance of viscous and yield stresses at the critical point, we employ a Bingham number defined as $Bi = \sigma_y/{\tau_v}$, where viscous stress  $\tau_v = \eta({\dot{\gamma})}\dot{\gamma}$. At the critical point, the Bingham number varies as $Bi \approx 0.9-0.97$ indicating that viscous stresses can barely deform the yield stress fluid around the cavity at the critical point. This result suggests that in experiments with the yield stress fluids, the medium around the cavity should experience a very weak deformation in the vicinity of the cavity and the elastic response of the surrounding material should be dominant. In fact, in fitting the experimental data of Fig.~\ref{Pd}(a,b) to Eq.~(\ref{eq_pcritical3_t}), we noticed that the pressure contribution from the plastically deformed zone near the surface of the cavity (indicated by $P_{out}|_1$) is much smaller than the elastic resistance of the surrounding fluid ($P_{out}|_1 << P_{out}|_2$). The latter result is consistent with the above hypothesis that the viscous stresses are not strong enough to significantly deform the yield stress fluids at the critical pressure and the elastic response of the yield stress material is dominant (i.e., ${\tau_v}/{E}<<1$ or ${\sigma_y}/{E}<<1$). }

\section{Conclusion}

In summary, we have performed needle induced cavitation experiments to assess the surface tension, Young modulus and the detailed form of flow structure around the cavity in a broad range of yield stress fluids. The findings of this study can be summarized as follows: First, the measured surface tension values for Carbopol based yield stress fluids is close \HMR{to surface tension of pure }water over a wide range of yield stress values ($\sigma_y = 0.5 -120$ Pa). Secondly, we demonstrated that NIC technique can be successfully used to measure the Young modulus as small as 10 Pa for yield stress gels. Thirdly, our flow visualization experiments revealed that \HMR{for $P\leq P_c$}, the fluid is barely perturbed by viscous stresses. However, post-instability the strength of the flow increases up to a local maximum in strain rate (or deformation rate) before it subsides gradually towards equilibrium at longer times. Finally, our results show that the flow strength post-instability is controlled by the critical pressure as well as the non-Newtonian rheology of the yield stress fluids. \par

Although we performed NIC experiments in a range of yield stress fluids with $\sigma_y = 0.5 -120$ Pa, by no means this method (and assessment of surface tension) is limited to this range of yield stress values and may be adopted for stiffer gels. Moreover, the impact of this work goes beyond the evaluation of the surface tension for yield stress materials. In fact, a broad range of biological tissues and cells are soft and assessing their Young modulus requires highly sensitive and sophisticated methods such as AFM or nano-indentation methods that often generates complex and ambiguity in the obtained results\cite{garcia2018,guz2014if}. The results of this work lend credence to NIC method as a fast and particularly straightforward technique that can be used for measuring the mechanical properties of biological samples\cite{pavlovsky2014}. This will be the focus of our future work.

\section{Supporting Information} 
Illustrates further information about the rheological properties of the fluids and the cavitation rheology experiments.

\section{Acknowledgement}
The authors are grateful to Anas Al-Humiari, Scott Hannahs and Richard Crisler for their help in implementing the NIC device. HM is grateful to Philipe Coussot, Randy Ewoldt and David Venerus for several helpful discussions.

\HMR{\section{Appendix: Plastic response of expanding bubble in a yield stress environment}
The calculation of the plastic response of an expanding bubble is based on the theoretical formulation by Bishop {\it et al.}\,\, \cite{bishop_theory_1945}.   We assume the surrounding material is extended between the internal radius of $R_0$ (here is the effective radius of the bubble as it starts to penetrate the surrounding elastic medium) and a large external radius of $R_{\infty}$ as shown in Fig.~\ref{Theo_Fig}a. The inner surface is under uniform pressure of $P_{out}$, while for a very large medium, we assume that there is no pressure on the external surface. 
Because of the spherical symmetry of the problem, a stress-based solution is used to formulate the problem  when the material remains elastic. The nonzero stress components are the radial stress $\sigma_r$ and the hoop stresses $\sigma_{\theta} = \sigma_{\phi}$. In the absence of body forces, the equilibrium equation in the radial direction is 
\begin{equation}
\frac{d \sigma_r}{dr}+\frac{2}{r}\left(\sigma_r-\sigma_{\theta}\right)\,=\,0.
\label{eq_rev0}
\end{equation}

In addition, the only displacement component is in the radial direction, $u=u(r)$ and the corresponding stain components in spherical polar coordinates are 
\begin{equation}
\epsilon_{rr} =\frac{d u}{dr},\hspace{1cm} \epsilon_{\theta\theta}\,=\,\epsilon_{\phi\phi}\,=\,\frac{u}{r}.
\label{eq_rev1}
\end{equation}

For a linearly elastic medium with Young Modulus of $E$ and Poisson ratio of$\nu$,  the stress-strain relations can be written as,
\begin{equation}
\epsilon_{rr}=\frac{1}{E} (\sigma_r -\nu \sigma_{\theta}), \hspace{1cm} \epsilon_{\theta\theta}=\frac{1}{E} \left[\sigma_{\theta} -\nu (\sigma_r+\sigma_{\theta})\right],
\label{eq_rev2}
\end{equation}
and to form Beltrami-Michell compatibility relationship of 
\begin{equation}
\frac{d}{dr}\left(\sigma_r+2\sigma_{\theta}\right)\,=\,0.
\label{eq_rev3}
\end{equation}

Eqs.\,\ref{eq_rev0} and \ref{eq_rev3} along with the boundary conditions can be solved  to find the Lame solution in spherical polar coordinate as \cite{asaro2006mechanics},
\begin{equation}
\sigma_r\,=\,-P_{out}\left(\frac{R_{\infty}^3}{r^3}-1\right)/\left(\frac{R_{\infty}^3}{R_0^3}-1\right) 
\label{eq_ve1a}
\end{equation}
\begin{equation}
\sigma_{\theta}\,=\,\sigma_{\phi}\,=\, P_{out}\left(\frac{R_{\infty}^3}{2 r^3}+1\right)/\left(\frac{R_{\infty}^3}{R_0^3}-1\right)
\label{eq_ve1b}
\end{equation}
where $r$ is the radial distance. Here, $\sigma_r$ is compressible stress and $\sigma_{\theta}$ is tensile stress and $|\sigma_r|>|\sigma_{\theta}|$. Equivalently, we can represent the spherical stress field  as the summation of the hydrostatic isotropic pressure of $-\sigma_{\theta}$ (equivalently a hydrostatic tension of $\sigma_{\theta}$) and uni-axial compressive stress in the radial direction and zero in other directions of $(\sigma_{\theta}-\sigma_r,0,0)$. This stress condition simplifies the von Mises  yield condition of the elastic materials as  \cite{hill1998mathematical}, 
\begin{equation}
\sigma_{\theta}-\sigma_r\,=\, \sigma_y
\label{eq_ve2}
\end{equation}

\begin{figure}[H]
\centering
	\includegraphics[width=0.7\textwidth]{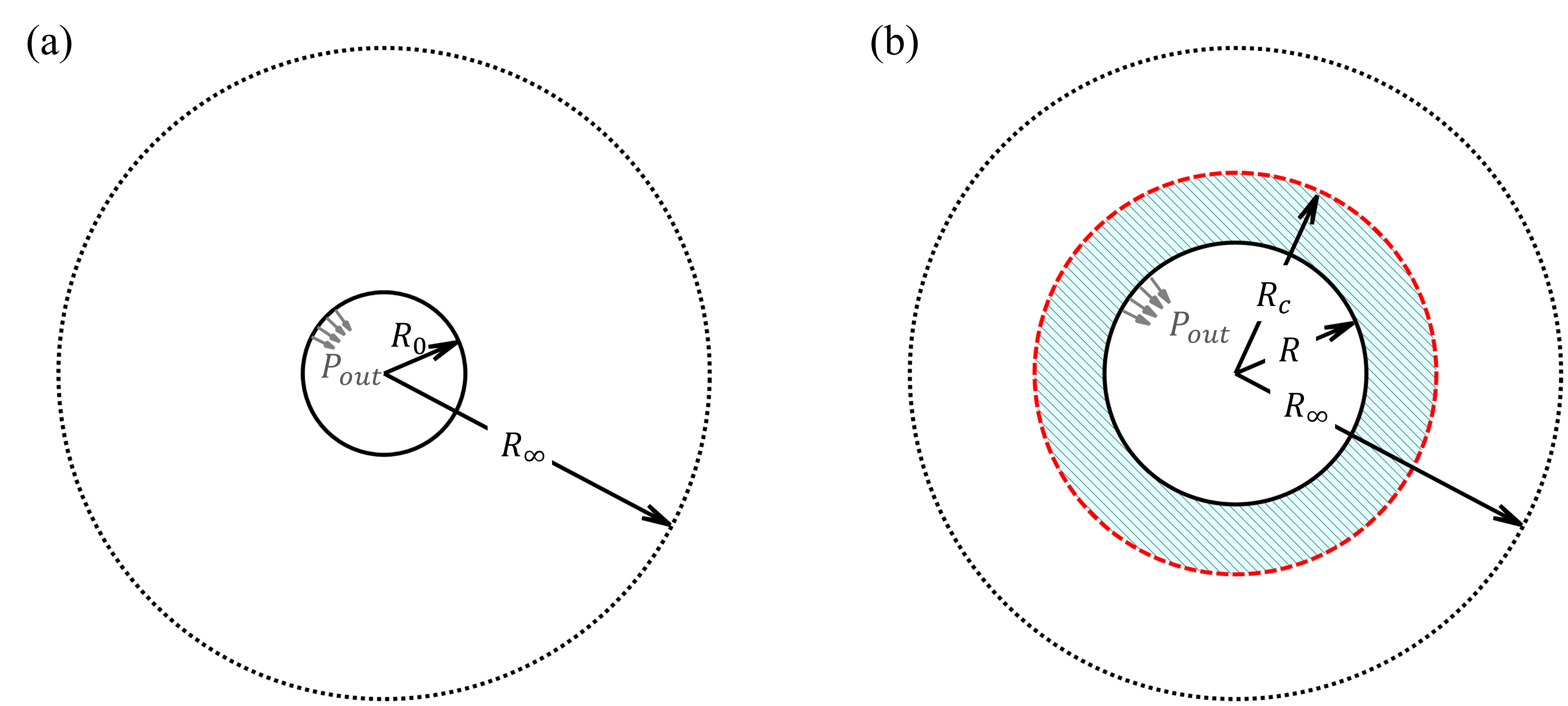}

	\caption{\small (a) Elastic region around a spherical bubble expanding by uniform distributed pressure $P_{out}$ at the initial stage. (b) The elastic and plastic regions around the expanded spherical bubble at a later stage.}\label{Theo_Fig}
\end{figure}

From Equations \ref{eq_ve1a},\ref{eq_ve1b} and \ref{eq_ve2}, we find that the corresponding pressure to the onset of the yield condition in linear elastic perfectly plastic materials is,
\begin{equation}
P_{out,\sigma_y}\,=\, \frac{2\sigma_y}{3} \left(1- \frac{R_0^3}{R_{\infty}^3}\right)
\label{eq_pcritical}
\end{equation}
For an infinite domain, we have $\frac{R_0}{R_{\infty}} \rightarrow 0$ and the inner layer of medium near the surface yields at a pressure equal to two-thirds of the yield stress. Given the typical $\sigma_y$ values of Carbopol gels, one can find that in almost all cases the material passes its purely elastic response regime very fast at the beginning of the bubble creation. Afterward, we have an elastic-perfectly plastic response. In this case, Eqs.\,\ref{eq_rev0} and \ref{eq_rev3} are used to find the stress components in the elastic region, while in the plastic region, the modified system of equations of Eqs.\,\ref{eq_rev0} and \ref{eq_ve2} are employed. To find the final solution, the continuation of the radial stress is used, and with the assumption that  the yield surface is placed at a radius $R_c$ (Fig. \ref{Theo_Fig}b), we can find the stress fields inside and outside the plastic regions as

\begin{minipage}{.4\linewidth}
\begin{equation}
\left.
              \begin{array}{ll}
              \sigma_r=-\frac{2\sigma_y}{3}\left(\frac{R_c^3}{r^3}\right)\\
              \sigma_{\theta}=\frac{2\sigma_y}{3}\left(\frac{R_c^3}{2r^3}\right)
            \end{array}
            \right\} \,\,\,\, c\leq r
\end{equation}
\end{minipage}%
\begin{minipage}{.55\linewidth}
\begin{equation}
\left.
              \begin{array}{ll}
              \sigma_r=-2\sigma_y \ln\left(\frac{R_c}{r}\right) -\frac{2\sigma_y}{3}\\
              \sigma_{\theta}=\sigma_y-2\sigma_y \ln\left(\frac{R_c}{r}\right) -\frac{2\sigma_y}{3}
            \end{array}
            \right\} \,\,\,\, c\geq r
\end{equation}
\end{minipage}

If the bubble is at the radius $R$, one can use the above relation and relate $R_c$ and $P_{out}$ as,
\begin{equation}
P_{out}\,=\,2 \sigma_y \ln\left(\frac{R_c}{R}\right)+\frac{2\sigma_y}{3}
\label{eq_pcritical}
\end{equation}

Unfortunately, the above relation is undetermined to provide a $P_{out}-R$ relation, as it involves an unknown $R_c$ value. To close the problem, a much more complicated solution procedure should be followed to calculate the elastic and plastic strain field based on $R$ and $R_c$ and to determine $R_c$ value.  A simplifying assumption can be made for the medium response by assuming that at a later time of bubble growth, the bubble behaves similarly to a spherical bubble that is expanded from zero radii. In this case, we can show that $R_c/R$ should remain constant based on the self-similarity and takes the form of \cite{hill1998mathematical}
\begin{equation}
\frac{R_c}{R}\,=\left(\frac{E}{3(1-\nu) \sigma_y}\right)^{\frac{1}{3}}
\label{eq_ca}
\end{equation}
and therefore we can rewrite the equation \ref{eq_pcritical} as 
\begin{equation}
P_{out}\,=\frac{2\sigma_y}{3} \left\{1+\ln\left(\frac{E}{3(1-\nu) \sigma_y}\right)\right\}
\label{eq_pcritical2}
\end{equation}

Assuming the $\nu$ of the incompressible medium is 0.5, we can find that $P_{out}$ is a constant independent of the bubble radius and is equal to 
\begin{equation}
P_{out}\,=\frac{2\sigma_y}{3} \left\{1+\ln\left(\frac{2E}{3\sigma_y}\right)\right\}
\label{eq_pcritical2}
\end{equation}
Given that we assume here the bubble expands from a zero initial radius, the above solution should be regarded as an upper limit of a bubble expansion from a finite-size needle in a perfect plastic regime. In the above model, we assumed that the stress level could not exceed a fixed yield limit. However, for many polymeric media, there is still elastic stress even after the yield condition \cite{n2019elastoplastic}, especially if the material is confined and can not undergo large permanent deformation. Equivalently, it is possible to define the terminal steady-state stress in the post-yield condition as \cite{hill1998mathematical} 
\begin{equation}\sigma\,=\, \sigma_y + h_{\sigma}(\epsilon)\label{eq_starinhardening}
\end{equation}
where $h_{\sigma}$ is a function expressing the change of total stress as a function of the logarithmic total strain $\epsilon$. Here, the material undergoes strain-hardening or work-hardening and the von-Mises yield condition of Eq. \ref{eq_ve2} is modified to \cite{hill1998mathematical},
\begin{equation}
\sigma_{\theta}-\sigma_r\,=\, \sigma_y\,+\, h \left\{2\ln\left(\frac{r}{r_0}\right)-\frac{1-2\nu}{E}\left(\sigma_r+\sigma_{\theta}\right) \right\}.
\end{equation}
where $r_0$ is the initial radial position of the element. 
If the medium is incompressible ($\nu=\frac{1}{2}$), the equilibrium condition of Eq. \ref{eq_rev0} can be simplified as \begin{equation}
\frac{d \sigma_r}{dr}\,=\,\frac{2}{r}\left(\sigma_{\theta}-\sigma_r\right)\,=\,\frac{2\sigma_y}{r}+\frac{2h_{\sigma}[2\ln(r/r_0)]}{r}
\label{eq_rev4}
\end{equation}

If we assume that the bubble is expanded from zero radius in the infinite medium, we can use the self-similarity property of the spherical shape and  scale spatial lengths based on the current size of the bubble $R$.
By adopting this condition, and solving  Eq. \ref{eq_rev4}, a a revised relation for $P_{out}$ can be found in the form of
\begin{equation}
P_{out}\,=\frac{2\sigma_y}{3} \left\{1+\ln\left(\frac{2E}{3\sigma_y}\right)\right\}+2\int_1^{\left(\frac{2E}{3\sigma_y}\right)^{\frac{1}{3}}}\,h_{\sigma}\left\{\frac{2}{3}\ln\left(\frac{t^3}{t^3-1}\right)\right\}\frac{dt}{t}
\label{eq_pcritical3}
\end{equation}
where for the linear $h_{\sigma}(\epsilon)=E \epsilon$ function (when the stress-strain curve is represented with a slope of $E$.) \cite{bishop_theory_1945}, the relation can be simplified as 
\begin{equation}
P_{out}\,=\frac{2\sigma_y}{3} \left\{1+\ln\left(\frac{2E}{3\sigma_y}\right)\right\}+\frac{2\pi^2}{27} E
\label{eq_pcritical3}
\end{equation}
}

\bibliographystyle{unsrt}
\bibliography{sample}

\begin{thebibliography}{10}

\bibitem{frigaard2019}
Ian Frigaard.
\newblock Simple yield stress fluids.
\newblock {\em Current Opinion in Colloid \& Interface Science}, 43:80--93,
  2019.

\bibitem{balmforth2014}
Neil~J Balmforth, Ian~A Frigaard, and Guillaume Ovarlez.
\newblock Yielding to stress: recent developments in viscoplastic fluid
  mechanics.
\newblock {\em Annu. Rev. Fluid Mech}, 46(1):121--146, 2014.

\bibitem{coussot2014}
Ph~Coussot.
\newblock Yield stress fluid flows: A review of experimental data.
\newblock {\em Journal of Non-Newtonian Fluid Mechanics}, 211:31--49, 2014.

\bibitem{malkin2017}
Alexander Malkin, Valery Kulichikhin, and Sergey Ilyin.
\newblock A modern look on yield stress fluids.
\newblock {\em Rheologica Acta}, 56(3):177--188, 2017.

\bibitem{saud2021}
Keara~T Saud, Mahesh Ganesan, and Michael~J Solomon.
\newblock Yield stress behavior of colloidal gels with embedded active
  particles.
\newblock {\em Journal of Rheology}, 65(2):225--239, 2021.

\bibitem{foudazi2015}
Reza Foudazi, Sahar Qavi, Irina Masalova, and Alexander~Ya Malkin.
\newblock Physical chemistry of highly concentrated emulsions.
\newblock {\em Advances in colloid and interface science}, 220:78--91, 2015.

\bibitem{ovarlez2010}
Guillaume Ovarlez, Quentin Barral, and Philippe Coussot.
\newblock Three-dimensional jamming and flows of soft glassy materials.
\newblock {\em Nature materials}, 9(2):115--119, 2010.

\bibitem{gross2014}
Markus Gross, Timm Kr{\"u}ger, and Fathollah Varnik.
\newblock Rheology of dense suspensions of elastic capsules: normal stresses,
  yield stress, jamming and confinement effects.
\newblock {\em Soft matter}, 10(24):4360--4372, 2014.

\bibitem{Bing22}
E.C. Bingham.
\newblock {\em Fluidity and plasticity}.
\newblock McGraw-Hill, New York, 1922.

\bibitem{nelson2020}
Arif~Z Nelson, Binu Kundukad, Wai~Kuan Wong, Saif~A Khan, and Patrick~S Doyle.
\newblock Embedded droplet printing in yield-stress fluids.
\newblock {\em Proceedings of the National Academy of Sciences},
  117(11):5671--5679, 2020.

\bibitem{SMALL2015}
Christina~C. Small, Sunny Cho, Zaher Hashisho, and Ania~C. Ulrich.
\newblock Emissions from oil sands tailings ponds: Review of tailings pond
  parameters and emission estimates.
\newblock {\em Journal of Petroleum Science and Engineering}, 127:490--501,
  2015.

\bibitem{pourzahedi2021}
A~Pourzahedi, M~Zare, and IA~Frigaard.
\newblock Eliminating injection and memory effects in bubble rise experiments
  within yield stress fluids.
\newblock {\em Journal of Non-Newtonian Fluid Mechanics}, 292:104531, 2021.

\bibitem{johnson2017}
Michael Johnson, Michael Fairweather, David Harbottle, Timothy~N Hunter,
  Jeffrey Peakall, and Simon Biggs.
\newblock Yield stress dependency on the evolution of bubble populations
  generated in consolidated soft sediments.
\newblock {\em AIChE Journal}, 63(9):3728--3742, 2017.

\bibitem{kam2001}
Seung~I Kam, Phillip~A Gauglitz, and William~R Rossen.
\newblock Effective compressibility of a bubbly slurry: I. theory of the
  behavior of bubbles trapped in porous media.
\newblock {\em Journal of colloid and interface science}, 241(1):248--259,
  2001.

\bibitem{israelachvili2011}
Jacob~N Israelachvili.
\newblock {\em Intermolecular and surface forces}.
\newblock Academic press, 2011.

\bibitem{HU1991723}
R.Y.Z. Hu, A.T.A. Wang, and J.P. Hartnett.
\newblock Surface tension measurement of aqueous polymer solutions.
\newblock {\em Experimental Thermal and Fluid Science}, 4(6):723--729, 1991.

\bibitem{MANGLIK200155}
Raj~M. Manglik, Vivek~M. Wasekar, and Juntao Zhang.
\newblock Dynamic and equilibrium surface tension of aqueous surfactant and
  polymeric solutions.
\newblock {\em Experimental Thermal and Fluid Science}, 25(1):55--64, 2001.

\bibitem{G_raud_2014}
Baudouin G{\'{e}}raud, Loren J{\o}rgensen, Laure Petit, H{\'{e}}l{\`{e}}ne
  Delanoë-Ayari, Pierre Jop, and Catherine Barentin.
\newblock Capillary rise of yield-stress fluids.
\newblock {\em {EPL} (Europhysics Letters)}, 107(5):58002, aug 2014.

\bibitem{Bouj13}
J.~Boujlel and P.~Coussot.
\newblock Measuring the surface tension of yield stress fluids.
\newblock {\em Soft Matter}, 9:5898--5908, 2013.

\bibitem{C5SM00569H}
Loren Jørgensen, Marie Le~Merrer, Hélène Delanoë-Ayari, and Catherine
  Barentin.
\newblock Yield stress and elasticity influence on surface tension
  measurements.
\newblock {\em Soft Matter}, 11:5111--5121, 2015.

\bibitem{Lopez18}
William~F. Lopez, Mônica~F. Naccache, and Paulo~R. de~Souza~Mendes.
\newblock Rising bubbles in yield stress materials.
\newblock {\em Journal of Rheology}, 62(1):209--219, 2018.

\bibitem{gutowski2012}
Iris~A Gutowski, David Lee, John~R de~Bruyn, and Barbara~J Frisken.
\newblock Scaling and mesostructure of carbopol dispersions.
\newblock {\em Rheologica acta}, 51(5):441--450, 2012.

\bibitem{jaworski2021}
Zdzis{\l}aw Jaworski, Tadeusz Spychaj, Anna Story, and Grzegorz Story.
\newblock Carbomer microgels as model yield-stress fluids.
\newblock {\em Reviews in Chemical Engineering}, 2021.

\bibitem{jalaal2015}
Maziyar Jalaal, Neil~J Balmforth, and Boris Stoeber.
\newblock Slip of spreading viscoplastic droplets.
\newblock {\em Langmuir}, 31(44):12071--12075, 2015.

\bibitem{walls2003}
HJ~Walls, S~Brett Caines, Angelica~M Sanchez, and Saad~A Khan.
\newblock Yield stress and wall slip phenomena in colloidal silica gels.
\newblock {\em Journal of Rheology}, 47(4):847--868, 2003.

\bibitem{bertola2003}
V~Bertola, F~Bertrand, H~Tabuteau, D~Bonn, and P~Coussot.
\newblock Wall slip and yielding in pasty materials.
\newblock {\em Journal of Rheology}, 47(5):1211--1226, 2003.

\bibitem{divoux2011}
Thibaut Divoux, Catherine Barentin, and S{\'e}bastien Manneville.
\newblock Stress overshoot in a simple yield stress fluid: An extensive study
  combining rheology and velocimetry.
\newblock {\em Soft Matter}, 7(19):9335--9349, 2011.

\bibitem{abbasi2021}
Aref Abbasi~Moud, Jade Poisson, Zachary~M Hudson, and Savvas~G Hatzikiriakos.
\newblock Yield stress and wall slip of kaolinite networks.
\newblock {\em Physics of Fluids}, 33(5):053105, 2021.

\bibitem{yildirim2005}
Ozgur~E Yildirim, Qi~Xu, and Osman~A Basaran.
\newblock Analysis of the drop weight method.
\newblock {\em Physics of Fluids}, 17(6):062107, 2005.

\bibitem{stauffer1965}
Clyde~E Stauffer.
\newblock The measurement of surface tension by the pendant drop technique.
\newblock {\em The journal of physical chemistry}, 69(6):1933--1938, 1965.

\bibitem{lecacheux2022}
Laure Lecacheux, Abdelkrim Sadoudi, Agn{\`e}s Duri, V{\'e}ronique Planchot, and
  Thierry Ruiz.
\newblock The role of laplace pressure in the maximal weight of pendant drops.
\newblock {\em Journal of Colloid and Interface Science}, 606:920--928, 2022.

\bibitem{mysels1990}
Karol~J Mysels.
\newblock The maximum bubble pressure method of measuring surface tension,
  revisited.
\newblock {\em Colloids and surfaces}, 43(2):241--262, 1990.

\bibitem{fainerman2004}
VB~Fainerman and R~Miller.
\newblock Maximum bubble pressure tensiometry—an analysis of experimental
  constraints.
\newblock {\em Advances in colloid and interface science}, 108:287--301, 2004.

\bibitem{simon1851}
M~Simon.
\newblock Recherches sur la capillarit{\'e}.
\newblock In {\em Annales de Chimie et de Physique}, volume~32, page~5, 1851.

\bibitem{kuffner1961}
Roy~J Kuffner.
\newblock The measurement of dynamic surface tensions of solutions of slowly
  diffusing molecules by the maximum bubble pressure method.
\newblock {\em Journal of Colloid Science}, 16(5):497--500, 1961.

\bibitem{austin1967}
M~Austin, BB~Bright, and EA~Simpson.
\newblock The measurement of the dynamic surface tension of manoxol ot
  solutions for freshly formed surfaces.
\newblock {\em Journal of Colloid and Interface Science}, 23(1):108--112, 1967.

\bibitem{joos1981}
P~Joos and E~Rillaerts.
\newblock Theory on the determination of the dynamic surface tension with the
  drop volume and maximum bubble pressure methods.
\newblock {\em Journal of Colloid and Interface Science}, 79(1):96--100, 1981.

\bibitem{mysels1986improvements}
Karol~J Mysels.
\newblock Improvements in the maximum-bubble-pressure method of measuring
  surface tension.
\newblock {\em Langmuir}, 2(4):428--432, 1986.

\bibitem{mysels1989some}
Karol~J Mysels.
\newblock Some limitations in the interpretation of the time dependence of
  surface tension measured by the maximum bubble pressure method.
\newblock {\em Langmuir}, 5(2):442--447, 1989.

\bibitem{fainerman1998maximum}
VB~Fainerman and R~Miller.
\newblock The maximum bubble pressure tensiometry.
\newblock In {\em Studies in Interface Science}, volume~6, pages 279--326.
  Elsevier, 1998.

\bibitem{sugden1924determination}
Samuel Sugden.
\newblock The determination of surface tension from the maximum pressure in
  bubbles.
\newblock {\em Journal of the Chemical Society}, 121:14, 1924.

\bibitem{bendure1971dynamic}
Raymond~L Bendure.
\newblock Dynamic surface tension determination with the maximum bubble
  pressure method.
\newblock {\em Journal of Colloid and Interface Science}, 35(2):238--248, 1971.

\bibitem{fainerman2004correction}
VB~Fainerman, VD~Mys, AV~Makievski, and R~Miller.
\newblock Correction for the aerodynamic resistance and viscosity in maximum
  bubble pressure tensiometry.
\newblock {\em Langmuir}, 20(5):1721--1723, 2004.

\bibitem{zimberlin2007}
Jessica~A Zimberlin, Naomi Sanabria-DeLong, Gregory~N Tew, and Alfred~J Crosby.
\newblock Cavitation rheology for soft materials.
\newblock {\em Soft Matter}, 3(6):763--767, 2007.

\bibitem{kundu2009}
Santanu Kundu and Alfred~J Crosby.
\newblock Cavitation and fracture behavior of polyacrylamide hydrogels.
\newblock {\em Soft Matter}, 5(20):3963--3968, 2009.

\bibitem{zimberlin2010}
Jessica~A Zimberlin, Jennifer~J McManus, and Alfred~J Crosby.
\newblock Cavitation rheology of the vitreous: mechanical properties of
  biological tissue.
\newblock {\em Soft Matter}, 6(15):3632--3635, 2010.

\bibitem{barney2020}
Christopher~W Barney, Carey~E Dougan, Kelly~R McLeod, Amir Kazemi-Moridani, Yue
  Zheng, Ziyu Ye, Sacchita Tiwari, Ipek Sacligil, Robert~A Riggleman,
  Shengqiang Cai, et~al.
\newblock Cavitation in soft matter.
\newblock {\em Proceedings of the National Academy of Sciences},
  117(17):9157--9165, 2020.

\bibitem{cui2011}
Jun Cui, Cheol~Hee Lee, Aline Delbos, Jennifer~J McManus, and Alfred~J Crosby.
\newblock Cavitation rheology of the eye lens.
\newblock {\em Soft Matter}, 7(17):7827--7831, 2011.

\bibitem{delbos2012}
Aline Delbos, Jun Cui, Sami Fakhouri, and Alfred~J Crosby.
\newblock Cavity growth in a triblock copolymer polymer gel.
\newblock {\em Soft Matter}, 8(31):8204--8208, 2012.

\bibitem{hutchens2016}
Shelby~B Hutchens, Sami Fakhouri, and Alfred~J Crosby.
\newblock Elastic cavitation and fracture via injection.
\newblock {\em Soft matter}, 12(9):2557--2566, 2016.

\bibitem{barney2019}
Christopher~W Barney, Yue Zheng, Shuai Wu, Shengqiang Cai, and Alfred~J Crosby.
\newblock Residual strain effects in needle-induced cavitation.
\newblock {\em Soft Matter}, 15(37):7390--7397, 2019.

\bibitem{gent1969}
AN~Gent and DA~Tompkins.
\newblock Nucleation and growth of gas bubbles in elastomers.
\newblock {\em Journal of applied physics}, 40(6):2520--2525, 1969.

\bibitem{canchi2017}
Saranya Canchi, Karen Kelly, Yu~Hong, Michael~A King, Ghatu Subhash, and Malisa
  Sarntinoranont.
\newblock Controlled single bubble cavitation collapse results in jet-induced
  injury in brain tissue.
\newblock {\em Journal of the mechanical behavior of biomedical materials},
  74:261--273, 2017.

\bibitem{Rassolov2020}
Peter Rassolov and Hadi Mohammadigoushki.
\newblock {Effects of elasticity and flow ramp up on kinetics of shear banding
  flow formation in wormlike micellar fluids}.
\newblock {\em Journal of Rheology}, 64:1161--1177, 2020.

\bibitem{Rassolov2022}
Peter Rassolov and Hadi Mohammadigoushki.
\newblock {Role of Micellar Entanglement Density on Kinetics of Shear Banding
  Flow Formation}.
\newblock {\em Journal of Rheology}, 67:169--181, 2023.

\bibitem{nazari2023helical}
Farshad Nazari, Kourosh Shoele, and Hadi Mohammadigoushki.
\newblock Helical locomotion in yield stress fluids.
\newblock {\em Physical Review Letters}, 130(11):114002, 2023.

\bibitem{Oel22}
Claude Oelschlaeger, Jonas Marten, Florian Péridont, and Norbert Willenbacher.
\newblock Imaging of the microstructure of carbopol dispersions and correlation
  with their macroelasticity: A micro- and macrorheological study.
\newblock {\em Journal of Rheology}, 66(4):749--760, 2022.

\bibitem{llewellin2002}
EW~Llewellin, HM~Mader, and SDR Wilson.
\newblock The rheology of a bubbly liquid.
\newblock {\em Proceedings of the Royal Society of London. Series A:
  Mathematical, Physical and Engineering Sciences}, 458(2020):987--1016, 2002.

\bibitem{borhan1999}
A~Borhan and J~Pallinti.
\newblock Breakup of drops and bubbles translating through cylindrical
  capillaries.
\newblock {\em Physics of Fluids}, 11(10):2846--2855, 1999.

\bibitem{garcia2018}
Pablo~D Garcia and Ricardo Garcia.
\newblock Determination of the elastic moduli of a single cell cultured on a
  rigid support by force microscopy.
\newblock {\em Biophysical journal}, 114(12):2923--2932, 2018.

\bibitem{guz2014if}
Nataliia Guz, Maxim Dokukin, Vivekanand Kalaparthi, and Igor Sokolov.
\newblock If cell mechanics can be described by elastic modulus: study of
  different models and probes used in indentation experiments.
\newblock {\em Biophysical journal}, 107(3):564--575, 2014.

\bibitem{pavlovsky2014}
Leonid Pavlovsky, Mahesh Ganesan, John~G Younger, and Michael~J Solomon.
\newblock Elasticity of microscale volumes of viscoelastic soft matter by
  cavitation rheometry.
\newblock {\em Applied Physics Letters}, 105(11):114105, 2014.

\bibitem{bishop_theory_1945}
R~F Bishop, R~Hill, and N~F Mott.
\newblock The theory of indentation and hardness tests.
\newblock {\em Proceedings of the Physical Society}, 57(3):147--159, May 1945.

\bibitem{asaro2006mechanics}
Robert Asaro and Vlado Lubarda.
\newblock {\em Mechanics of solids and materials}.
\newblock Cambridge University Press, 2006.

\bibitem{hill1998mathematical}
Rodney Hill.
\newblock {\em The mathematical theory of plasticity}, volume~11.
\newblock Oxford university press, 1998.

\bibitem{n2019elastoplastic}
E~N’gouamba, Julie Goyon, and Philippe Coussot.
\newblock Elastoplastic behavior of yield stress fluids.
\newblock {\em Physical Review Fluids}, 4(12):123301, 2019.

\end{thebibliography}

\end{document}


\title{This file serves as supplementary information for the paper: Cavitation Rheology of Model Yield Stress Fluids Based on Carbopol}

	\author{Hadi Mohammadigoushki}
	\email[]{hadi.moham@eng.famu.fsu.edu}
	\affiliation{Department of Chemical and Biomedical Engineering, FAMU-FSU College of Engineering, Tallahassee, FL 32310, USA.}

 \author{Kourosh Shoele}
\affiliation{Department of Mechanical Engineering, FAMU-FSU College of Engineering, Florida State University, Tallahassee, FL, 32310, USA}
\date{\today}

\maketitle

\renewcommand{\thefigure}{S\arabic{figure}}
\renewcommand{\thetable}{S\arabic{table}}

\begin{table*}[h!]
\begin{tabular}{c c c c c c}

\hline\hline

 Fluid & Carbopol concentration [wt$\%$]&$\sigma_y$[Pa] & $K$ [Pa.s$^{n}$] & n & $G_o$  [Pa] \\ 
 \hline
 
 pure corn syrup & -- & 0 & 9.8 & 1 & -- \\ 
 \hline 

\multirow{11}{2cm}{\centering 940} 
&0.025  & ---   & $2\times 10^{-3}$        & 1& ---  \\  
& 0.05  & ---   &$3\times 10^{-2}$         & 0.89& ---  \\
& 0.06  & ---   & $3.8\times 10^{-1}$      & 0.52 & ---  \\
& 0.065 & ---   & $7.2\times 10^{-1}$      & 0.46&---  \\
& 0.07  & 0.9   & 1.18                     & 0.41& 2.8  \\
& 0.075 & 2   & 1.92        & 0.44 & 30  \\
& 0.085 & 4.4   & 3.1        & 0.38 & 39  \\
& 0.1 & 16   & 10.35        & 0.36& 144  \\
& 0.15 & 46   & 17.6        & 0.31 & 320  \\
& 0.2 & 69   & 33.5       & 0.27& 400  \\
& 0.5 & 120   & 57.1        & 0.27& 650  \\

\hline
\multirow{11}{2cm}{\centering Ultrez-10}& 
0.04 & ---  & $7.4\times 10^{-3}$        & 0.9 & --- \\ 
& 0.05 & ---   & $3\times 10^{-1}$       & 0.75& ---  \\
& 0.06 & 0.55   & 1        & 0.43& 4.5  \\
& 0.07 & 4   & 4.3       & 0.39& 38  \\
& 0.08 & 6.5   & 5.2        & 0.39& 55  \\
& 0.1 & 14.5   & 10.3       & 0.37& 130  \\
& 0.15 & 39   & 24.1      & 0.33& 260  \\
& 0.2 & 46   & 30.3       & 0.29& 299  \\
& 0.3 & 64   & 46.8   & 0.28        &  390  \\
& 0.4 & 85   & 55.8   & 0.29        & 480  \\
& 0.5 & 95  & 70   & 0.27  & 560  \\
 \hline
\end{tabular}
\caption{A summary of the rheological properties of all fluids used in this study along with the best fits to the Herschel-Bulkley results.  }
\label{table:2}
\end{table*}
 
\begin{figure}
     \centering
     \includegraphics[width=01\textwidth]{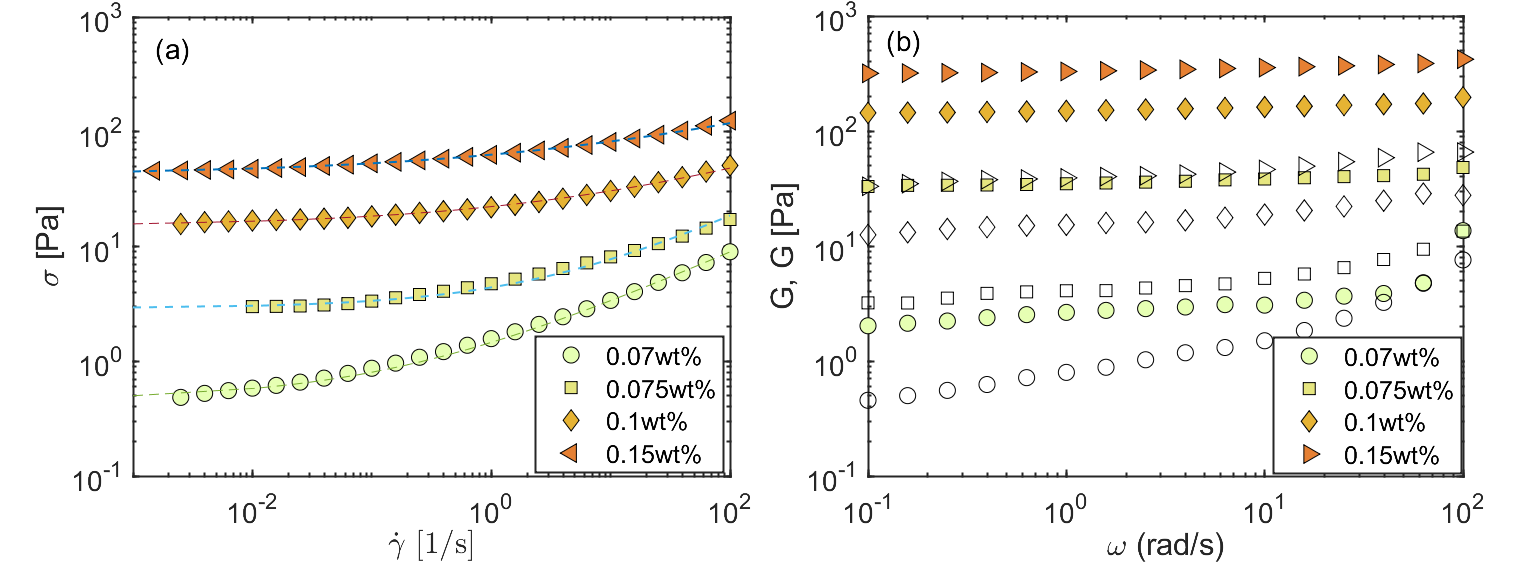}
 \caption{(a) steady shear stress as a function of applied shear rate for yield stress fluids based on Carbopol 940. The dashed curves show the best fit to the Herschel-Bulkley fluid model. (b) Storage (filled symbols) and loss moduli (open symbols) as a function of angular frequency.}
     \label{fig:rheolog_940}
     
\end{figure}

\begin{figure}
     \centering
     \includegraphics[width=0.6\textwidth]{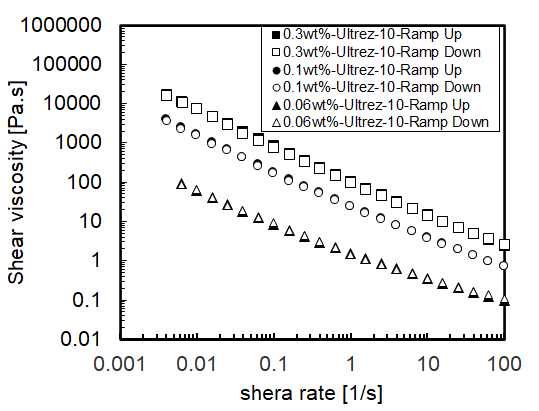}
 \caption{(a) steady shear viscosity as a function of applied shear rate for yield stress fluids based on Carbopol Ultrez-10. These experiments were performed by increasing the applied shear rate (Ramp-Up; closed symbols) and also by decreasing the applied shear rate (Ramp-Down; open symbols).}
     \label{fig:rheolog_940}
     
\end{figure}   
   
\begin{figure}[H]
\centering
\vspace{-5cm}
 \setlength{\unitlength}{1\textwidth}
\begin{picture}(1,1)
    \put(0,0){\includegraphics[width=1\textwidth]{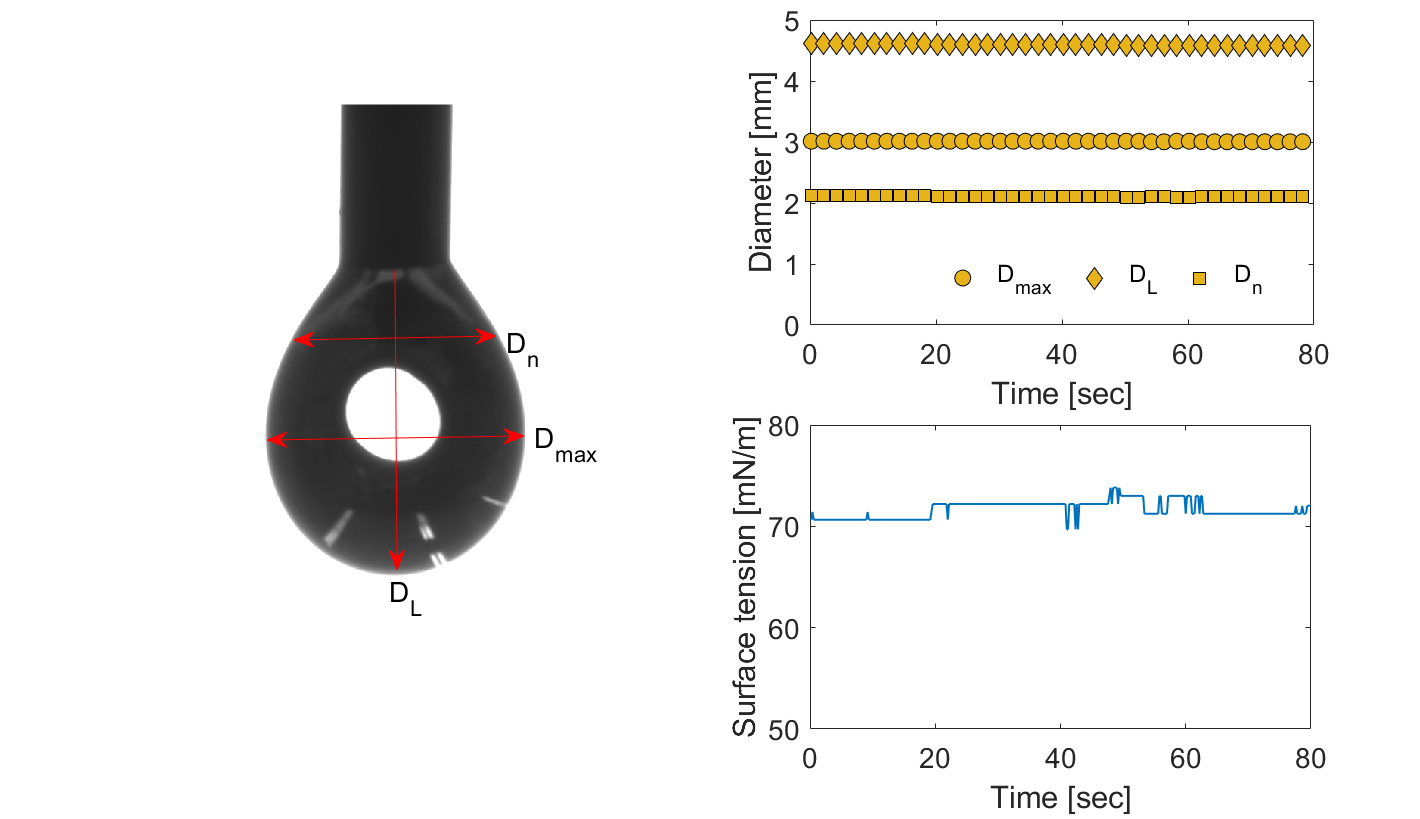}}
      \figlabel[0.3]{0.01}{0.5}{(a)}
     \figlabel[0.4]{0.3}{0.55}{(b)}
      \figlabel[0.4]{0.3}{0.3}{(c)}
    
     
    \end{picture}
	\caption{\small (a) A snapshot of the pendant droplet of a Carbopol solution (940; 0.025 wt$\%$) that is formed below a capillary tube along with important geometrical features that are used for calculation of the surface tension. Here $D_{max}$, $D_{L}$ and $D_{n}$ refer to the maximum diameter of the droplet, length of the droplet and the diameter of the droplet neck, respectively. To measure each of these dimensions, we developed our MATLAB script that works based on the edge detection algorithm. (b) The change in geometrical parameter of the pendant drop as a function of time. (c) Calculated surface tension of this system as a function of time. To estimate the surface tension using pendent drop analysis, we used a proposed procedure which is detailed in: Rusanov, A.I.; Prokhorov, V.A. Interfacial Tensiometry; Elsevier: Amsterdam, 1996 as well as in: Stauffer, C.E. The measurement of surface tension by the pendent drop technique. J. Phys. Chem. 1965, 69, 1933-1938.
 }\label{Pendant_drop}
\end{figure}
  \begin{figure}[H]
\centering
	\includegraphics[width=0.5\textwidth]{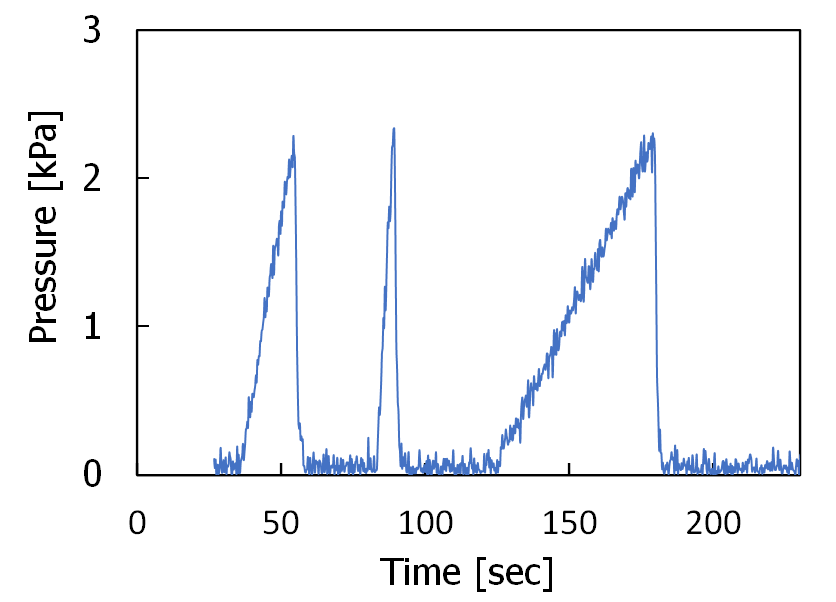}

	\caption{\small Pressure as a function of time for a needle with d = 0.31 mm inside a 0.5 wt$\%$ Carbopol Ultrez-10 solution. The inject rates from left to right are 1, 10 and 0.3 $\mu$L/hr, respectively.}\label{pressure_rate}
\end{figure}

	